\documentclass[11pt,a4paper]{article}

\pdfoutput=1
\usepackage{jcappub}
\usepackage{blindtext}
\usepackage{hyperref}
\usepackage{amsmath,amssymb}
\usepackage{float}
\usepackage{microtype}

\usepackage{graphicx}
\usepackage{bm}
\usepackage{latexsym}
\usepackage{epsfig}
\usepackage{psfrag}
\usepackage{color}
\usepackage[dvipsnames]{xcolor}
\usepackage{subfigure}
\usepackage{graphicx}

\usepackage{amssymb}
\usepackage{amsmath}
\usepackage{bm}
\usepackage{latexsym}
\usepackage{epsfig}
\usepackage{psfrag}
\usepackage{amsmath}
\usepackage[normalem]{ulem}
\usepackage{textcomp}
\usepackage{color}
\usepackage{pstricks}
\usepackage[utf8]{inputenc}
\usepackage{array}
\usepackage{comment}
\usepackage{mathalpha}

\makeatletter
\gdef\@fpheader{}
\makeatother

\def\bea{\begin{eqnarray}}
\def\eea{\end{eqnarray}}
\def\nn{\nonumber}

\def\f{\frac}
\def\l{\left}
\def\r{\right}
\def\d{{\rm d}}
\def\Mpl{M_{_{\mathrm{Pl}}}}
\def\mpcinv{{\rm Mpc}^{-1}}

\def\ps{\mathcal{P}_{_{\mathrm{S}}}}
\def\pc{\mathcal{P}_{_{\mathrm{C}}}}

\def\pcq{\mathcal{P}^{(4)}_{_{\mathrm{C}}}}

\def\cAs{\mathcal{A}_{_{\mathrm{S}}}}

\def\ei{\eta_{\rm i}}

\def\ee{\eta_{\rm e}}

\def\fnl{f_{_{\rm NL}}}

\def\vk{{\bm{k}}}

\def\d{{\mathrm{d}}}

\allowdisplaybreaks

\begin{document}
\title{Loop contributions to the scalar power spectrum due to quartic order 
action in ultra slow roll inflation}
\author{Suvashis Maity$^{1\,+}$,}
\emailAdd{$^+$suvashis@physics.iitm.ac.in}
\author{H.~V.~Ragavendra$^{2\,\ast}$,}
\emailAdd{$^\ast$ragavendra@rrimail.rri.res.in}
\author{Shiv K. Sethi$^{2\,\dagger}$ and}
\emailAdd{$^{\dagger}$sethi@rri.res.in}
\author{L.~Sriramkumar$^{1\,\ddagger}$}
\emailAdd{$^\ddagger$sriram@physics.iitm.ac.in}
\affiliation{$^1$Centre for Strings, Gravitation, and Cosmology, Department of 
Physics, Indian Institute of Technology Madras, Chennai 600036, India}
\affiliation{$^2$Raman Research Institute, C.~V.~Raman Avenue, Sadashivanagar, 
Bengaluru~560080, India}

\abstract
{The investigation of the theory of inflation beyond the linear
order in perturbations is important both for theoretical consistency and 
potential observables. 
In the contemporary literature, the calculation of modifications to the 
inflationary scalar power spectrum due to the loops from the higher order
interaction terms in the Hamiltonian have led to an interesting discussion 
regarding the validity of perturbation theory and the robustness of its 
predictions. 
Recently, there have been many efforts to examine the contributions to the 
scalar power spectrum due to the loops arising from the cubic order terms 
in the action describing the perturbations, specifically in inflationary 
scenarios that permit an epoch of ultra slow roll~(USR). 
A brief phase of USR during inflation is known to lead to interesting features 
in the scalar power spectrum which in turn has significant observational
consequences, such as the copious production of primordial black holes. 
In this work, we consider the loop contributions to the scalar power spectrum 
in a scenario of USR inflation arising {\it due to the quartic order terms}\/ 
in the action describing the scalar perturbations.
We compute the loop contributions to the scalar power spectrum due to the 
dominant term in the action at the quartic order in a scenario wherein a 
short phase of USR is sandwiched between two stages of slow roll (SR) inflation. 
We analyze the behaviour of the loop contributions in terms of the parameters 
that characterize the non-trivial inflationary dynamics, viz. the onset and 
duration of USR, and the smoothness of transitions between the USR and SR phases. 
We examine three different cases of the scenario---the late, intermediate and
early epochs of USR during inflation, each of which affects the scalar power
spectrum over different ranges of wave numbers. 
In the inflationary scenario involving a late phase of USR, for reasonable choices 
of the parameters, we show that the loop corrections are negligible for the entire 
range of wave numbers.
In the intermediate case, the contributions from the loops prove to be scale 
invariant over large scales and, we find that these contributions can amount
to $30\%$ of the leading order (i.e. the Gaussian) power spectrum. 
In the case wherein USR sets in early, we find that the loop contributions could 
be negative and can dominate the power spectrum at the leading order, which
indicates a breakdown of the validity of the perturbative expansion.  
We discuss the origin of the negative sign and the divergences that arise in the
loop contributions to the power spectrum.
We conclude with a brief summary and outlook.}

\maketitle


\section{Introduction}

The observational predictions of the inflationary paradigm are broadly in 
agreement with the cosmological data at the zeroth and first orders in 
perturbation theory (see, for example, the reviews~\cite{1992PhR...215..203M,
Martin:2003bt,Martin:2004um,Bassett:2005xm,Sriramkumar:2009kg,Baumann:2008bn,
Baumann:2009ds,Sriramkumar:2012mik,Linde:2014nna,Martin:2015dha}).
In particular, the generic predictions of the slow roll inflationary scenario
driven by a canonical scalar field at the leading and linear orders in 
perturbation theory---such as a spatially flat universe and a spectral 
index of scalar perturbations that is close to, but smaller than unity---are
remarkably consistent with the data from Planck on the anisotropies in the 
cosmic microwave background (CMB)~\cite{Ade:2015lrj,Planck:2018jri}.
There are ongoing efforts to observe the direct imprints of primary 
gravitational waves (GWs) (i.e. the tensor perturbations generated 
from the vacuum during inflation) through the so-called B-mode 
polarization of the CMB~\cite{LiteBIRD:2020khw}\footnote{In this regard, see, 
for example, https://www.isas.jaxa.jp/en/missions/spacecraft/future/litebird.html.}.
If these signatures are detected, they would immediately help in determining
the energy scale of inflation.  
In such a situation, it has become imperative to inquire whether the theory 
of inflation is well behaved and whether it makes testable predictions at the 
higher orders in perturbation theory. 

Over the last two decades, there has been a significant effort in the
literature to understand the observational predictions beyond the linear
order in perturbation theory (i.e. at the cubic and higher orders in the 
actions governing the perturbations), in particular, to examine the extent 
of non-Gaussianities generated during inflation. 
With the cubic and higher order actions in hand, the non-Gaussianities are 
often calculated using the standard methods of perturbative quantum field 
theory (in this context, see, for instance, Refs.~\cite{Maldacena:2002vr,
Seery:2005wm,Seery:2006wk,Chen:2010xka,Chen:2006dfn}).
As is well known in quantum field theory, cubic and higher order terms in 
an action describing a field can lead to loop corrections to the two-point 
correlation functions determined by the original, quadratic action.
In a similar manner, the higher order terms in the action describing the 
cosmological perturbations can lead to modifications to the power spectra
evaluated at the quadratic order (or, equivalently, in the linear theory).
In fact, the calculations of such higher order corrections to the inflationary
power spectra provide an opportunity to examine the validity of the perturbative 
expansion~\cite{Weinberg:2005vy,Weinberg:2006ac,Sloth:2006az,Sloth:2006nu,
Seery:2007wf,Byrnes:2007tm,Adshead:2008gk,Cogollo:2008bi,Rodriguez:2008hy,
Dimastrogiovanni:2008af,Senatore:2009cf,Bartolo:2010bu,Weinberg:2010wq}. 
Such an exercise has garnered renewed attention during the last year or two
particularly in the context of models of inflation admitting a brief epoch 
of ultra slow roll~(USR), which lead to an enhancement of the scalar power 
over small scales and hence produce significant population of primordial 
black holes~(PBHs)~\cite{Syu:2019uwx,Kristiano:2021urj,Inomata:2022yte,
Kristiano:2022maq,Riotto:2023hoz,Kristiano:2023scm,
Choudhury:2023vuj,Choudhury:2023jlt,Choudhury:2023rks,
Firouzjahi:2023aum,Firouzjahi:2023ahg,Franciolini:2023lgy,
Tasinato:2023ukp,Cheng:2023ikq,Fumagalli:2023hpa}.
We should also mention that there has been several efforts in the literature 
to evaluate the non-Gaussianities generated in such scenarios and understand 
their implications for observational probes such as the spectral density of
GWs and the scalar power spectrum in the dark ages observed through the~$21\,
\mathrm{cm}$ line emitted by neutral hydrogen~(in this context, see, for
example, Refs.~\cite{Unal:2018yaa,Cai:2018dig,Ragavendra:2020sop,Adshead:2021hnm,
Ragavendra:2021qdu,Balaji:2022zur,Balaji:2022rsy,Garcia-Saenz:2022tzu,Chen:2022dah,Ota:2022xni,Yamauchi:2022fri}).

It seems reasonable to expect that, in the slow roll (SR) inflationary scenario,
the higher order terms in the Hamiltonian do not lead to significant loop
corrections to the scalar power spectrum.
The corrections in SR inflation typically involve higher powers of 
dimensionless scalar power spectrum, whose amplitude is constrained to be 
about $\cAs \simeq 2\times 10^{-9}$ by the CMB data. 
Since the scalar power spectrum is nearly the same over all scales, the flow 
of power from small to large scales through the non-linear terms in the action
is reflected as a logarithmic factor in the contributions due to the loops, 
which leads to negligible corrections (in this regard, see Refs.~\cite{Weinberg:2005vy,
Weinberg:2006ac,Sloth:2006az,Sloth:2006nu,Seery:2007wf,Byrnes:2007tm,
Adshead:2008gk,Cogollo:2008bi,Rodriguez:2008hy,Dimastrogiovanni:2008af,
Senatore:2009cf,Bartolo:2010bu,Weinberg:2010wq}).
However, in inflationary models with a phase of USR, the scalar power can 
rise by seven to eight orders of magnitude on small scales (for a recent
review on the topic, see Ref.~\cite{Ragavendra:2023ret}). 
In such situations, it is conceivable that the higher order contributions 
from loops become comparable to the quantities arrived at from the linear
theory, thus violating the validity of perturbative expansion (for the 
ongoing debate on this issue, see Refs.~\cite{Kristiano:2022maq,Riotto:2023hoz,
Kristiano:2023scm,Choudhury:2023vuj,Firouzjahi:2023aum,Firouzjahi:2023ahg,
Franciolini:2023lgy}).

The discussion surrounding this important issue has largely remained focused 
on the contributions to the power spectrum due to the loops arising from the 
action governing the curvature perturbation at the cubic order, though there 
have also been some efforts to investigate the corresponding effects from the 
action at the quartic order~\cite{Ota:2022xni,Firouzjahi:2023aum}.
In the calculation of the contributions due to the loops, different studies 
in the literature have adopted varying methods. 
One approach relies on using the interaction picture in quantum field theory
(for details, see, for example, Ref.~\cite{Weinberg:2005vy}). 
The interaction Hamiltonian is first arrived at the desired order in 
perturbation theory.
The contributions due to the loops is given by the expectation value of the
two-point correlation function of the scalar perturbations, now elevated to 
operators, due to the interaction Hamiltonian~(in this context, see, for 
instance, Refs.~\cite{Kristiano:2022maq,Kristiano:2023scm}).
The other method is the classical treatment of the scalar perturbations, where 
one solves their equation of motion, containing a source term arising due to 
the higher order terms in the action (as discussed in Refs.~\cite{Kristiano:2023scm,
Riotto:2023hoz}).
Another method that has been adopted involves the calculation of the scalar 
non-Gaussianity parameter, say, $\fnl$, associated with the higher order 
action and computing the modifications to the spectrum that arises due to the
redefinition of the perturbation in terms of the parameter (as adopted in
Refs.~\cite{Unal:2018yaa,Cai:2018dig,Adshead:2021hnm,Ragavendra:2021qdu,Das:2023cum}).
Yet another method relies on computing the modification of the Friedmann equations 
due to the non-trivial dynamics of the model and accounting for them in the equation 
of motion of the scalar perturbations~\cite{Boyanovsky:2005px,
Boyanovsky:2005sh,Syu:2019uwx,Cheng:2021lif,Cheng:2023ikq}.
While the equivalence of the first two methods have been explicitly shown and
the third has been discussed to some extent in the literature, the equivalence 
and the implications of all these methods remains to be examined carefully.

In this work, we shall evaluate the loop contributions to the scalar power spectrum
arising due to the dominant term in the action at the quartic order that describes 
the scalar perturbations. 
To carry out the calculation, we shall adopt the first of the methods mentioned above.
In the approach, the scalar perturbations are treated as quantum operators and 
the loop contributions to the spectrum due to the quartic order terms are 
evaluated in the interaction picture.
We should highlight the fact that, as the field is Gaussian at the linear order, 
the contributions to the power spectrum due to the quartic order action {\it precedes}\/ 
the contributions due to the cubic order action.
The reason being that, while the former involves a six-point function of the 
perturbation, the latter requires one to compute the eight-point function. 
Moreover the quartic order action captures more transparently the non-trivial 
aspects of model when there are deviations from the slow roll dynamics.
This is due to the presence of slow roll parameters that are higher in order than 
those in the cubic order action. 
Hence, studying this contribution is essential to draw conclusions regarding 
the validity of the perturbative expansion of the system and the observational 
implications of the loop contributions.

We consider an inflationary scenario wherein a brief epoch of USR is 
sandwiched between two SR regimes. 
We model the different phases of inflation using the behavior of the first 
SR parameter. 
Further, we model the transition between the SR and USR phases consistently 
through the higher order SR parameters, which allows us to explore cases of 
instantaneous and smooth transitions. 
We calculate the loop contributions due to the dominant term in the quartic 
order action in such a scenario and analyse its behavior with respect to the 
parameters that determine the nature of the USR epoch, viz. its onset, its
duration, and the smoothness of the transition to and from the phase of~USR.

An important issue in the calculation of loop contributions to the standard,
scalar power spectrum at the linear order is the occurrence of the 
infrared~(IR) and the ultraviolet~(UV) divergences. 
Various studies, in the past and recently, have discussed the manner in which
these divergences are to be handled~\cite{Collins:2005nu,Burgess:2009bs,
Seery:2010kh,Gerstenlauer:2011ti,Tasinato:2023ukp}.
As we shall see, the IR divergence arises in the super Hubble limit and it
is logarithmic in form. 
Whereas, the UV divergence that occurs in the sub Hubble limit is quadratic 
in nature.
We find that the latter poses a more serious challenge than the former 
and we shall discuss them when we encounter them in the calculations.

This paper is organized as follows. 
In the next section, we shall describe the three-phase model of interest, 
as parametrized by the slow roll parameters.
and the behaviour of the Fourier
modes of the scalar 
perturbations in the different regimes. 
In Sec.~\ref{sec:loops}, we shall present the structure of the dominant 
contributions to the scalar power spectrum due to the loops arising from 
the action at the quartic order.
In Sec.~\ref{sec:results}, we shall discuss the dependence of this 
contribution on the parameters that describe the epoch of USR. 
We shall discuss the effects due to early and late onset of USR separately
and the associated observational implications.
In Sec.~\ref{sec:conc}, we shall summarize our results and present a brief
outlook.

At this stage of our discussion, let us clarify a few points concerning
the conventions and notations that we shall work with.
In this paper, we shall work with natural units such that $\hbar=c=1$ and 
set the reduced Planck mass to be $\Mpl=\l(8\,\pi\, G\r)^{-1/2}$.
We shall adopt the signature of the metric to be~$(-,+,+,+)$.
We shall assume the background to be the spatially flat
Friedmann-Lema\^itre-Robertson-Walker~(FLRW) 
line element described by the scale factor~$a$ and the Hubble parameter~$H$.
Also, an overdot and an overprime shall denote differentiation with respect 
to the cosmic time~$t$ and the conformal time~$\eta$, respectively.


\section{Model of interest and the power spectrum at the linear 
order}\label{sec:model}

Our starting point is the following differential equation that governs the 
Mukhanov-Sasaki variable that describes the scalar perturbations~\cite{1992PhR...215..203M}:
\begin{equation}
v_k''+ \l(k^2 - \f{z''}{z}\r)\, v_k = 0.
\end{equation}
The solutions to the equation satisfy the Wronskian condition 
\begin{equation}
v'^{*} v_k- v'_k v_k^* = i,
\label{eq:quanorm}
\end{equation}
where an overprime denotes derivative with respect to conformal time $\eta=\int dt/a$.
$z = a\Mpl\sqrt{2\epsilon_1}$
and $\epsilon_1 = H'\eta/H$ is the first slow-rollover parameter.
The quantity $z''/z$ can be expressed in terms of  slow-roll parameters as~\cite{Martin:2011sn}
\begin{eqnarray}
\f{z''}{z} &=& a^2\,H^2\,\l(2-\epsilon_1+\f{3\,\epsilon_2}{2}
+\f{\epsilon_2^2}{4}-\f{\epsilon_1\,\epsilon_2}{2}
+\f{\epsilon_2\,\epsilon_3}{2}\r).\label{eq:zppbz}
\end{eqnarray}
The higher order slow roll parameters  $\epsilon_i$  for $i \ge 2$ are defined as $\epsilon_{i}= \d \ln \epsilon_{i-1}/\d N
\simeq -\eta\,\d \ln \epsilon_{i-1}/\d \eta$ where  $N$ denotes e-folds. While all  the slow-roll parameters are small
for slow roll inflation, some of the parameters could assume large values during the USR phase of inflation, e.g.  $\epsilon_2 = -6$ during the USR phase. 

We are interested in an  inflationary scenario in which a brief USR phase is sandwiched between two SR phases.   
We model these  three phases  in terms of the evolution of the  first slow roll parameter $\epsilon_1$:
\begin{equation}
\epsilon_1(\eta) =
\begin{cases}
\epsilon_{1{\rm i}} 
& \text{in phase I with $\eta < \eta_1$}, \\
\epsilon_{1{\rm i}}\,\l({\eta/\eta_1}\r)^6
& \text{in phase II with $\eta_1 < \eta < \eta_2$}, \\
\epsilon_{1{\rm f}}
&\text{in phase III with $\eta > \eta_2$},
\end{cases}
\end{equation}
where $\epsilon_{1{\mathrm f}}=\epsilon_{1{\mathrm i}}\,(\eta_2/\eta_1)^6$.
Evidently, the parameters $\eta_1$ and $\eta_2$ denote the conformal times at 
the beginning and the end of the USR phase. While such scenario can be realized with an inflationary potential with an inflection point, it is not our aim here to consider a specific model but rather to compute the observables in terms of the smoothness and the duration of the transition. The  relative duration of the three phases will be discussed and justified in a later section. 

It can be shown that the last term of the RHS of Eq.~(\ref{eq:zppbz}) ($\propto \epsilon_3$) dominates $z''/z$ during  the 
SR-USR-SR transitions.
One may expect the term such as $\epsilon_1\epsilon_2$ to dominate at the transitions,
due to the abrupt change in the behaviour of $\epsilon_1$ at these points. However,
the value of $\epsilon_1$ in SR and USR regimes is much smaller than unity. On the 
other hand, the value of $\epsilon_2$, though much lesser than unity in SR, becomes
$-6$ in the USR regime. Thus the change in $\epsilon_2$ is much more pronounced than
that of $\epsilon_1$. So, the quantity $\epsilon_2\epsilon_3 \sim \epsilon_2'$
dominates over the earlier term. Such behaviours of these slow roll parameters have
been consistently seen in models of USR in the literature (see, for example,
Ref.~\cite{Ragavendra:2020sop}).
Assuming an instantaneous transition, we get (for the first transition):
\begin{equation}
\f{z''}{z} 
\simeq \f{a\,H\,\epsilon_2'}{2} 
= -\f{\epsilon_2^{_{\mathrm{II}}}-\epsilon_2^{_{\mathrm{I}}}}{2\,\eta}\,
\delta^{(1)}(\eta-\eta_1).\label{eq:zppz-dd}
\end{equation}
It is our aim to study the impact of the smoothness of this transition on cosmological observables. This motivates us to 
model $\epsilon_2'$ as: 
\begin{equation}
\epsilon_2' = 
\begin{cases}
\f{\epsilon_2^{\rm II}}{\sqrt{\pi}\Delta\eta}\,
\mathrm{e}^{-\f{(\eta-\eta_{1})^2}{\Delta\eta^2}}~~~~~\text{around $\eta_1$}, \\
-\f{\epsilon_2^{\rm II}}{\sqrt{\pi}\Delta\eta}\,
\mathrm{e}^{-\f{(\eta-\eta_{2})^2}{\Delta\eta^2}}~~~\text{around $\eta_2$},
\end{cases}
\end{equation}
Here we have used  $\epsilon_2^{\rm I, III} \ll 1$ (during the  SR phases) and $\epsilon_2^{\rm II}=-6$. 
The parameter $\Delta \eta$ dictates the smoothness of the transition. 
In the limit $\Delta \eta \to 0$, 
$\epsilon_2' \to \pm \epsilon^{\rm II}_2\delta^{(1)}(\eta-\eta_{1,2})$.
Using this characterization of transition, we determine the higher order slow 
roll parameters to be
\begin{subequations}\label{eq:epss}
\begin{eqnarray}
\epsilon_3 &=& \mp \f{\eta}{\sqrt{\pi}\Delta \eta}
\mathrm{e}^{-\f{(\eta-\eta_{1,2})^2}{\Delta\eta^2}},\\
\epsilon_4 &=& -\left[{1-2\f{\eta(\eta-\eta_{1,2})}{\Delta\eta^2}}\right]\,,\\
\epsilon_5 &=& 2\f{\eta(2\eta-\eta_{1,2})}{\Delta\eta^2-2\eta(\eta-\eta_{1,2})}.
\end{eqnarray}
\end{subequations}
We note that $\epsilon_3$ contains the parametric form of the Dirac delta 
function with which we modelled $\epsilon_2'$. Further, $\epsilon_4$ contains
a term $\propto 1/\Delta\eta^2$ that diverges strongly in the limit 
$\Delta \eta \to 0$.
However, $\epsilon_5$ tends to a finite value in the same limit. These
behaviours shall be crucial in identifying the relevant terms that dominate
the action at the quartic order and hence the associated correction to the scalar
power spectrum.
Having modelled the slow roll parameters, we can compute any quantity 
describing the background dynamics, provided we also specify the value of
Hubble parameter $H_{\rm I}$ at an initial time\footnote{Note that $H_{\rm I}$,
though is the value of the parameter at an initial time, can be taken as its 
value throughout the evolution, since the value of 
$\epsilon_1 \equiv -\dot H/H^2$ is $< 1$ if not $\ll 1$ at any given time.}.
This value is determined  by  the observed CMB angular power 
spectrum~\cite{Aghanim:2018eyx,Planck:2018jri}. 

We next discuss the solutions of Mukhanov-Sasaki equation during the three 
inflationary phases. 
During the initial phase of slow roll,  i.e. during $\eta < \eta_1$, the 
solution to the Mukhanov-Sasaki equation with the Bunch-Davies initial 
condition and normalization  (Eq.~(\ref{eq:quanorm})), is given by
\begin{equation}
v_k^{_{\mathrm{I}}}(\eta)
=\f{1}{\sqrt{2\,k}}\,
\l(1-\f{i}{k\,\eta}\r)\, \mathrm{e}^{-i\,k\,\eta}.
\label{eq:vk-I}
\end{equation}

During the USR phase,  $\eta_2>\eta>\eta_1$, the solution to the 
Mukhanov-Sasaki equation  is:
\begin{eqnarray}
v_k^{_{\mathrm{II}}}(\eta)
&=& \f{\gamma_k}{\sqrt{2\,k}}\,
\l(1-\f{i}{k\,\eta}\r)\,  \mathrm{e}^{-i\,k\,\eta}
+ \f{\delta_k}{\sqrt{2\,k}}\,
\l(1+\f{i}{k\,\eta}\r)\,\mathrm{e}^{i\,k\,\eta}
\label{eq:vk-II}
\end{eqnarray}
The coefficients~$\gamma_k$ and $\delta_k$ are determined by the matching 
conditions on the Mukhanov-Sasaki variable~$v_k$ and its derivative
$v_k'$ at~$\eta_1$ as:
\begin{subequations}\label{eq:mc2}
\begin{eqnarray}
v_k^{_{\mathrm{II}}}(\eta_1) &=& v_k^{_{\mathrm{I}}}(\eta_1),\\
v_k^{_{\mathrm{II}}\prime}(\eta_1) 
&=& v_k^{_{\mathrm{I}}\prime}(\eta_1)
-  \f{\epsilon_2^{_{\mathrm{II}}}-\epsilon_2^{_{\mathrm{I}}}}{2\,\eta_1}\,
v_k^{_{\mathrm{I}}}(\eta_1).
\end{eqnarray}
\end{subequations}
On satisfying these condition, we obtain:
\begin{subequations}\label{eq:ab-sr-usr-p6}
\begin{eqnarray}
\gamma_k &=& 1+\f{3\,i}{2\,k\,\eta_1}+\f{3\,i}{2\,k^3\,\eta_1^3},\\
\delta_k &=& \l(-\f{3\, i}{2\, k\,\eta_1} 
- \frac{3}{k^2\,\eta_1^2} +\frac{3\, i}{2\, k^3\,\eta_1^3}\r)\,
\mathrm{e}^{-2\, i\, k\,\eta_1},
\end{eqnarray}
\end{subequations}

In the third regime when $\eta > \eta_2$, since it is a slow roll phase again, 
the solution for~$v_k$ can be expressed as
\begin{eqnarray}
v_k^{_{\mathrm{III}}}(\eta)
&=&\f{\alpha_k}{\sqrt{2\,k}}\,
\l(1-\f{i}{k\,\eta}\r)\,\mathrm{e}^{-i\,k\,\eta}
+ \f{\beta_k}{\sqrt{2\,k}}\, \l(1+\f{i}{k\,\eta}\r)\, 
\mathrm{e}^{i\,k\,\eta}.
\label{eq:vk-III}
\end{eqnarray} 
Since the transition at $\eta_2$ will again lead to a Dirac delta function 
in the quantity $z''/z$  (as in Eq.~\eqref{eq:zppz-dd}, but with the 
corresponding values of $\epsilon_2$ across the transition), we can obtain the 
coefficients~$\alpha_k$ and~$\beta_k$ by matching the solutions and their 
derivatives at~$\eta_2$ using the conditions
\begin{subequations}\label{eq:mc2}
\begin{eqnarray}
v_k^{_{\mathrm{III}}}(\eta_2) &=& v_k^{_{\mathrm{II}}}(\eta_2),\\
v_k^{_{\mathrm{III}}\prime}(\eta_2) 
&=& v_k^{_{\mathrm{II}}\prime}(\eta_2)
-  \f{\epsilon_2^{_{\mathrm{III}}}-\epsilon_2^{_{\mathrm{II}}}}{2\,\eta_2}\,
v_k^{_{\mathrm{II}}}(\eta_2).
\end{eqnarray}
\end{subequations}
We find that the quantities~$\alpha_k$ and $\beta_k$ are given by
\begin{subequations}
\begin{eqnarray}
\alpha_k 
&=& \l(1 - \f{3\,i}{2\,k\,\eta_2}
-\f{3\,i}{2\,k^3\,\eta_2^3}\r)\, \gamma_k
- \l(\f{3\,i}{2\,k\,\eta_2}-\f{3}{k^2\,\eta_2^2}
-\f{3\,i}{2\,k^3\,\eta_2^3}\r)\,
\delta_k\,\mathrm{e}^{2\,i\,k\,\eta_2},\nn \\ \\
\beta_k 
&=& \l(1 + \f{3\,i}{2\,k\,\eta_2}
+\f{3\,i}{2\,k^3\,\eta_2^3}\r)\, \delta_k
+ \l(\f{3\,i}{2\,k\,\eta_2}+\f{3}{k^2\,\eta_2^2}
-\f{3\,i}{2\,k^3\,\eta_2^3}\r)\,
\gamma_k\,\mathrm{e}^{-2\,i\,k\,\eta_2}.\nn \\
\end{eqnarray} 
\end{subequations}
The Mukhanov-Sasaki variable is related to the gauge-invariant curvature perturbation $\zeta$ as $f_k(\eta)=v_k(\eta)/z(\eta)$, where $f_k$ is the mode function associated with $\zeta$ (for details of physical interpretation of $\zeta$ see e.g. \cite{1992PhR...215..203M,dodelson}).
The first-order power spectrum of $\zeta$ can then be evaluated in the final slow roll 
regime using the above results for the coefficients~$\alpha_k$ and~$\beta_k$:
\begin{eqnarray}
\ps(k) \equiv \f{k^3}{2\pi^2} \vert f_k \vert_{\eta \to 0}^2 &=& 
\f{H_{\rm I}^2}{8\pi^2\Mpl^2\,\epsilon_{1_{\rm f}}}\vert \alpha_k - \beta_k \vert^2\,.
\label{eq:powspec1}
\end{eqnarray}
We do not explicitly write down the complete expression for this tree-level 
power spectrum as it is rather long and cumbersome. However, we may understand
the behaviour of $\ps(k)$ by considering its asymptotic forms. 

Let us define $k_1 = -1/\eta_1$ and $k_2 = -1/\eta_2$, the wavenumbers that cross the Hubble radius at the time of two transitions. On large scales with 
$k \ll k_1 < k_2$, Eq.~(\ref{eq:powspec1}) reduces to 
\begin{eqnarray}
\ps(k) & \simeq & \f{H_{\rm I}^2}{8\pi^2\Mpl^2\,\epsilon_{1_{\rm i}}}\,,
\label{eq:ps-large-k}
\end{eqnarray}
and over small scales with $k \gg k_2 > k_1$, it assumes the form
\begin{eqnarray}
\ps(k) & \simeq & \f{H_{\rm I}^2}{8\pi^2\Mpl^2\,\epsilon_{1_{\rm f}}}\,.
\label{eq:ps-small-k}
\end{eqnarray}
These are clearly scale invariant regimes, only differing in amplitudes due to
the values of $\epsilon_1$ in the respective epochs of slow roll when these 
modes exit the Hubble radius. These behaviours suggest that the scales that are
far removed from $k_1$ and $k_2$ are largely unaffected by the dynamics of USR
except for the change in the value of $\epsilon_1$.
The scales that are intermediate to these regimes, that exit the Hubble radius 
during and close to USR, see a sharp change in the shape of the spectrum.
The spectrum contains a null at $k_{\rm dip} \simeq \sqrt{3}\,k_1\exp(-3\Delta N/2)$
where $\Delta N$ is the duration of USR measured in 
e-folds~\cite{Goswami:2010qu,Ozsoy:2021pws,Balaji:2022zur}.
The null is followed by a steep rise of $k^4$ until it peaks in its value and
gradually settles to the scale invariant form over smaller scales~\cite{Byrnes:2018txb,Ragavendra:2020sop,Tasinato:2020vdk,Cole:2022xqc}.
The spectrum contains tiny oscillations following the peak and persisting over
the small scales. Such oscillations arise mainly due to the instantaneous nature
assumed for the transitions and are generally absent in realistic scenarios of
USR epochs arising from a smooth potential driving inflation.
We should mention that we shall ignore this dependence of tiny oscillations on
the smoothness of the transition $\Delta \eta$, since we are mainly interested
in the overall amplitude and general shape of $\ps(k)$ and subsequently the
loop-level spectrum.
We shall illustrate the behaviour of this $\ps(k)$ when we arrive at the loop-level
contribution to it from the quartic action and compare them against each other.


\section{Calculation of the loop contributions due to the dominant term in
the action at the quartic order}\label{sec:loops}

We then turn to examine the action governing the scalar perturbation at the
quartic order and compute the corresponding loop-level contribution to 
$\ps(k)$.  Quartic order action has been considered  by many authors (e.g. \cite{Seery:2006vu, Chen:2006dfn,Jarnhus:2007ia,Dimastrogiovanni:2008af,
Bartolo:2010bu} partial list). In this paper, we study  the fourth order action computed by Dimastrogiovanni et~al. (2008) \cite{Dimastrogiovanni:2008af} in spatially flat gauge. It is our aim to  select the terms that lead to dominant contribution due to the
sharp nature of the transition between slow roll and USR phases.  Based on the discussion in the last section, these are terms
that contain higher order  slow roll parameters and lead to   highest powers in  the parameter characterizing  the sharpness of transitions, $1/\Delta \eta$.
After careful scrutiny of all the terms in the fourth order action,  we identify the following term as the most dominant contributor  at the quartic order from the Eq.~[20] of Ref.~\cite{Dimastrogiovanni:2008af}
\bea
\delta {\cal S}_4[\delta \phi] = -\f{1}{24}\int \d t \, \int\d^{3} {\bm x} \, 
a^3 V_{\phi\phi\phi\phi} \delta \phi^4\,,
\eea
where $V_{\phi\phi\phi\phi}=\d^4 V/\d\phi^4$.
Amongst the several terms present in the action, we choose this term using the 
following logic.
We express  all the relevant expressions  in terms of  the slow roll parameters and try to identify
the terms that contain highest order of these parameters. Since we are interested 
in a model with sharp transition between slow roll and USR phases, we are motivated 
to look for terms that contain $\epsilon_2'$ or its higher derivatives in time 
that lead to large contributions at the transition.
This method  of identifying the terms that  dominate  at the 
transitions of USR epochs is along the lines of similar analyses at 
the level of cubic order action (e.g. \cite{Kristiano:2022maq} and 
subsequent works). In these works, the term identified as the dominant one contains 
$\epsilon_2'$ which is a sharply peaking function at the transition. At the quartic 
order, the situation is more severe as there are higher order derivatives of 
$\epsilon_2'$. The dominant term in our case contains terms 
such as $\epsilon_2'''$, which we  recast in terms of higher order slow roll
parameters.

This procedure of identifying the dominant terms yields: 
\bea
\delta {\cal S}_{4}[\cal \delta \phi] &=& \f{1}{288\Mpl^4} \int 
\d \eta \, \int \d^{3}{\bm x} \, a^4 V \, \f{\delta \phi^4 }{\epsilon_1}
\Big[
\epsilon_2\epsilon_3\epsilon_4\epsilon_5
+3\epsilon_2\epsilon_3^2\epsilon_4
+\f{1}{2}\epsilon_2^2\epsilon_3\epsilon_4
+3\epsilon_2\epsilon_3\epsilon_4
-9\epsilon_1\epsilon_2\epsilon_3\epsilon_4\nn\\
&& +\epsilon_2 \epsilon_3 \epsilon_4^2
+\epsilon_2\epsilon_3^3
+\f{3}{2}\epsilon_2^2\epsilon_3^2
+3 \epsilon_2\epsilon_3^2 
-9 \epsilon_1\epsilon_2\epsilon_3^2
-\f{1}{2}\epsilon_2^3\epsilon_3
-\f{3}{2}\epsilon_2^2\epsilon_3
-\f{35}{2}\epsilon_1\epsilon_2^2\epsilon_3
+32\epsilon_1^2\epsilon_2\epsilon_3\nn\\
&& -24 \epsilon_1\epsilon_2\epsilon_3
-3\epsilon_1\epsilon_2^3  +39\epsilon_1^2 \epsilon_2^2
-9 \epsilon_1\epsilon_2^2  -56\epsilon_1^3\epsilon_2
+72 \epsilon_1^2\epsilon_2 +16 \epsilon_1^4 -48\epsilon_1^3
\Big].
\eea
Of the above terms we should note that the first term with 
$\epsilon_2\epsilon_3\epsilon_4\epsilon_5$ contains 
$\delta^{(1)}(\eta-\eta_{1,2})/\Delta\eta^2$. Other terms containing higher powers
of $\epsilon_3$ or $\epsilon_4$ may appear to lead to stronger divergence as
$\Delta\eta \to 0$. But they have the following behaviours and caveats.
The term with $\epsilon_3\epsilon_4^2$ in the limit $\Delta\eta \to 0$
behave as
\bea
\epsilon_3\,\epsilon_4^2 &\simeq & \eta\,\delta^{(1)}(\eta-\eta_{1,2})
\left[1-2\f{\eta(\eta-\eta_{1,2})}{\Delta\eta^2}\right]^2\,.
\eea
In the integrals where such terms appear, upon utilizing the Dirac delta function,
they effectively become
\bea
\epsilon_3\,\epsilon_4^2 &\simeq& 4\f{\eta_{1,2}^3\delta\eta^2}{\Delta\eta^4}
\delta^{(1)}(\eta-\eta_{1,2}) + {\cal O}\left(\f{1}{\Delta\eta^2}\right)\,,
\eea
where $\delta\eta \equiv \eta-\eta_{1,2}$. If $\delta\eta=\Delta\eta$ and
$\Delta \eta \to 0$ then
\bea
\epsilon_3\epsilon_4^2 &\simeq & 4\f{\eta_{1,2}^3}{\Delta\eta^2} 
\delta^{(1)}(\eta-\eta_{1,2}) + {\cal O}\left(\f{1}{\Delta\eta}\right)\,.
\eea
On the other hand, the terms with $\epsilon_3^2\epsilon_4$ behave as
\bea
\epsilon_3^2\epsilon_4
&\simeq & 2\,\delta^{(1)}(\eta-\eta_{1,2})\f{\eta^2\,\mathrm{e}^{\f{-(\eta-\eta_{1,2})^2}{\Delta\eta^2}}}{\Delta\eta}
\f{\eta(\eta-\eta_{1,2})}{\Delta\eta^2} + {\cal O}\left(\f{1}{\Delta\eta^2}\right)\,.
\eea
Once again, due to the Dirac delta function, integration involving these terms
lead to
\bea
\epsilon_3^2\epsilon_4 &\simeq& 2\,\eta^3_{1,2}\f{\mathrm{e}^{\f{-\delta\eta^2}{\Delta\eta^2}}}{\Delta\eta}
\f{\delta\eta}{\Delta\eta^2}\delta^{(1)}(\eta-\eta_{1,2})
+ {\cal O}\left(\f{1}{\Delta\eta^2}\right), \\
&\simeq& 2\f{\eta^3_{1,2}}{\Delta\eta^2}\delta^{(1)}(\eta-\eta_{1,2}) 
+ {\cal O}\left(\f{1}{\Delta\eta}\right)\,,
\eea
as $\delta \eta = \Delta \eta \to 0$. Lastly, the terms with $\epsilon_3\epsilon_4$
behave as
\bea
\epsilon_3\epsilon_4 &\simeq & 2\f{\eta^2\delta \eta}{\Delta\eta^2}\delta^{(1)}(\eta-\eta_{1,2})
 + {\cal O}\left(\f{1}{\Delta\eta}\right)\,, \\
&\simeq & 2\f{\eta^2_{1,2}}{\Delta\eta}\delta^{(1)}(\eta-\eta_{1,2})
+ {\cal O}\left(\f{1}{\Delta\eta^0}\right)\,,
\eea
using the same argument as above.
Hence, the terms with $\epsilon_3,\,\epsilon_4$ or their combinations with higher 
powers, apart from the first term lead to divergence similar to the first term 
($\sim 1/\Delta\eta^2$) or lesser in strength.
However, these terms arise from $V_{\phi\phi\phi\phi}$ as well as other lower order 
derivatives of potential. So, these terms arising from different derivatives may 
get added up or cancelled with one another.
This is not true for the first term as no lower derivative of the potential
shall lead to a term containing $\epsilon_5$. 
Besides, there can be boundary terms in the quartic order action that may contain
derivatives of the potential and hence slow-roll parameters that are comparable to the 
term of our interest. However, we can easily show that they shall not lead to any loop 
contribution at the same order as the the term of interest. We discuss this possibility 
in some length in App.~\ref{app:boundary}.
Therefore, we can see that the first term with $\epsilon_2\epsilon_3\epsilon_4\epsilon_5$ 
is unique and leads to the strongest possible divergence in terms of $\Delta\eta$ at the 
level of quartic action. Hence, we shall focus on this term which is a good 
representative of the dominant term of quartic order action and examine the 
characteristics of the corresponding loop-level contribution to the power spectrum.

We next  express the quartic action in terms of 
$\zeta$,  the gauge-invariant variable  related to  cosmological  observables ($\zeta = -3/2 \psi$ 
in the radiation dominated era at super-Hubble scales where $\psi$ is the Newtonian potential, e.g. \cite{dodelson}). 
In the  spatially flat 
gauge in which the quartic action has been computed \cite{Dimastrogiovanni:2008af}, $\zeta$ is related to 
the scalar field fluctuation $\delta \phi$ as (for switching between uniform density and spatially flat gauge and relation between relevant variables see e.g. \cite{Maldacena:2002vr}):
\begin{equation}
\zeta = \f{H}{\dot \phi}\delta\phi = -\f{\delta \phi}{\Mpl\sqrt{2\epsilon_1}}\,.
\label{eq:zeta-delphi}
\end{equation}
Eq.~(\ref{eq:zeta-delphi}) allows us to express the  quartic action in terms of
$\zeta$ as
\begin{eqnarray}
\delta {\cal S}_4[\zeta] &\supset& \f{1}{72}\int \d \eta \, \int \d^3{\bm x} \, 
a^4 V \left(\epsilon_1\epsilon_2\epsilon_3\epsilon_4\epsilon_5\right) 
\zeta^4\,.
\end{eqnarray}
The corresponding interaction Hamiltonian is given by\footnote{
$H^{(4)}_{\rm int}=\pi_\zeta \dot \zeta-{\cal L}^{(4)}$ where $\pi_\zeta$ is the
conjugate momentum of $\zeta$~[cf.~Eq.~14 in Ref.~\cite{Chen:2006dfn}]. 
But the term of our interest arises from the Lagrangian and a similar term 
$\sim \zeta^4$ does not arise from $\pi_\zeta \dot\zeta$. So the term of interest in 
$H^{(4)}_{\rm int}$ is simply the same as in the action but with a negative sign.}
\begin{eqnarray}
H^{(4)}_{\rm int} &\supset& -\f{1}{72}\int \d^3\bm x \, a^4 V 
(\epsilon_1\epsilon_2\epsilon_3\epsilon_4\epsilon_5) \zeta^4(\eta,\bm x)\,.
\label{eq:Hint}
\end{eqnarray}
Given the interaction Hamiltonian we can  compute the correction to  the power spectrum $\ps(k)$  using in-in formalism (for details of the in-in formalism and its detailed derivation see \cite{2005PhRvD..72d3514W}). We denote the correction  to the power spectrum
from the quartic interaction Hamiltonian as $\pc^{(4)}(k)$. We consider the two-point correlation of $\zeta$ in 
Fourier space and calculate the correction due to the term of interest in 
$H^{(4)}_{\rm int}$ at the leading order as
\begin{eqnarray}
\langle 0 \vert \hat \zeta_{\bm k}(\ee) \hat \zeta_{{\bm k}'}(\ee) \vert 0 \rangle 
& \simeq & 
\langle 0 \vert \hat \zeta_{\bm k}(\ee) \hat \zeta_{{\bm k}'}(\ee) \vert 0 \rangle
- i \langle 0 \vert \left[ \hat \zeta_{\bm k}(\ee) \hat \zeta_{{\bm k}'}(\ee)\,,
\int \d \eta \,{\cal T} \left( \hat H^{(4)}_{\rm int}(\eta, \bm x) \right) \right]\vert 0 \rangle. \nn \\
\label{eq:ps+pc}
\end{eqnarray}
Note that the perturbations are now treated as quantum operators denoted as $\hat \zeta_{\bm k}$. 
The corresponding mode functions $f_k(\eta)$ are related to them as
\bea
\hat \zeta_{\bm k}(\eta) &=& f_k(\eta)\hat a_{\bm k} \mathrm{e}^{i \vk \cdot {\bm x}} + 
f_k^\ast(\eta)\hat a^\dag_{\bm k} \mathrm{e}^{-i \vk \cdot {\bm x}}\,,
\eea
where $a_{\bm k}$ and $a^\dag_{\bm k}$ are the annihilation and creation operators 
corresponding to the mode ${\bm k}$ and $f_k$ is the associated mode function.
The second term of Eq.~\eqref{eq:ps+pc}, along with the explicit form of 
$H^{(4)}_{\rm int}$, gives the correction to the two-point function of the form
\begin{eqnarray}
\langle 0 \vert \hat \zeta_{\bm k} \hat \zeta_{{\bm k}'}(\ee) \vert 0 \rangle_{_{\rm C}} 
& \simeq & \f{i}{6}\,f_k(\ee) f_{k'}(\ee) \delta^{(3)}(\vk+\vk')
\int \d \eta\, a^4 V \epsilon_1\epsilon_2\epsilon_3\epsilon_4\epsilon_5
f^\ast_k(\eta)f^\ast_{k'}(\eta)
\int \f{\d^3 \bm q}{(2\pi)^3} \vert f_q(\eta) \vert^2 \nn \\
& & +~\text{complex conjugate}\,.
\label{eq:zeta-zeta-correction}
\end{eqnarray}
Note that it is a six-point correlation that has been reduced by using Wick's theorem
 to this form. As mentioned earlier, this contribution precedes the eight-point
correlation that arises from cubic order action. 
(For details about the contractions and associated Feynman diagrams of this 
contribution, see the  Appendix~\ref{app:contrac}.)
The above relation leads to the corresponding dimensionless power spectrum, which 
we shall denote as $\pcq(k)$, of the form
\begin{eqnarray}
\pcq(k) & = & \f{i}{6}\left(\f{k^3}{2\pi^2}\right)\,f_k^2(\ee)
\int \d \eta\, a^4 V \epsilon_1\epsilon_2\epsilon_3\epsilon_4\epsilon_5
\left[f^\ast_k(\eta)\right]^2 \int \d \ln q \, \ps(q,\eta) \nn \\
& & +~\text{complex conjugate}\,,
\label{eq:pc-def}
\end{eqnarray}
where we have used the definition of dimensionless power spectrum, $\ps(k,\eta)$:
\begin{equation}
\ps(k,\eta) = \f{k^3}{2\pi^2}\vert f_k(\eta) \vert^2.
\label{eq:ps-def}
\end{equation}
This $\pcq(k)$ is the contribution to $\ps(k)$ due to quartic order action and
the complete power spectrum at this order shall be $\ps(k)+\pcq(k)$.
\subsection{Slow roll}
Before computing $\pcq(k)$ in the SR-USR-SR model, we estimate the magnitude of 
such a loop-level contribution for the case of slow roll inflation.
In a standard scenario of slow roll, the slow roll parameters can be shown to be 
$\epsilon_1 \simeq \epsilon_2 \simeq \epsilon_3 \simeq \epsilon_4 \simeq \epsilon_5 \simeq 10^{-3} $. Further, we
can solve for the mode function $f_k(\eta)$ analytically and impose the 
Bunch-Davies initial condition when the mode is deep inside the Hubble radius. 
The mode function $f_k(\eta)$ can be expressed  as (e.g. \cite{1992PhR...215..203M,Maldacena:2002vr,Martin:2011sn,Ragavendra:2023ret} and references therein)
\begin{equation}
f_k(\eta) = -\f{H\eta}{\Mpl\sqrt{4k\epsilon_1}} \mathrm{e}^{-ik\eta}\left( 1 - \f{i}{k\eta}\right).
\end{equation}
Using this mode function and the the slow roll parameters  $\epsilon_i$s in Eq.~\eqref{eq:pc-def} we 
evaluate $\pcq(k)$ to be
\begin{eqnarray}
\pcq(k) &=& -\f{i}{8} 
\epsilon_2 \epsilon_3 \epsilon_4 \epsilon_5 
\left(\ps^0\right)^2\,\int \f{\d \eta}{k\,\eta^2} 
\left( 1 + \f{i}{k\eta} \right)^2 \mathrm{e}^{2ik\eta}
\int \d \ln q\, (1 + q^2\eta^2) \nn \\
& &+~\text{complex conjugate},
\end{eqnarray}
where $\ps^0 = H^2/(8\pi^2\Mpl^2\epsilon_1)$; in slow roll inflation, $\ps^0$ yields the first-order power spectrum 
(Eq.~(\ref{eq:ps-large-k})).
We have used the slow roll 
approximation that $3H^2\Mpl^2 \simeq V$ and $a = -1/(H\eta)$ in this relation. \
On performing the integral over $q$, we obtain
\begin{eqnarray}
\pcq(k) &=& -\f{i}{8} \epsilon_1^4
\left(\ps^0\right)^2\,\int^{\ee}_{\ei} \f{\d \eta}{k\,\eta^2} 
\left( 1 + \f{i}{k\eta} \right)^2 \mathrm{e}^{2ik\eta}
\bigg[\ln \left( \f{k_{\rm max}}{k_{\rm min}} \right) 
+ \f{\eta^2}{2}\left( k_{\rm max}^2 - k_{\rm min}^2 \right)\bigg] \nn \\
& &+~\text{complex conjugate}\,.
\label{eq:intoq}
\end{eqnarray}
We have introduced and used $k_{\rm min}$ and $k_{\rm max}$ as the limits of the 
integral over $q$, instead of taking the formal limits of $0$ and $\infty$, 
respectively. These quantities help us to illustrate the functional behaviour of 
the divergences that occur $\pcq(k)$ in these limits.
The quadratic term  $k_{\rm max}^2 - k_{\rm min}^2$ 
arises   owing  to the sub-Hubble behaviour of the 
mode function, whereas the super-Hubble behaviour of the mode function leads to the logarithmic term 
$\ln(k_{\rm max}/k_{\rm min})$.
We shall encounter these two types of divergences arising from integrating over 
$\ps(q,\eta)$ once again in the model of SR-USR-SR and shall discuss them 
separately in App.~\ref{app:divergence} in some detail.

To obtain numerical estimates, we set  $k_{\rm min} =10^{-6}\,\mpcinv$, 
the wavenumber that exits the Hubble radius 70 e-folds before the end of inflation. We choose
 $k_{\rm max} = 10^{20}\,\mpcinv$, $k_{\rm max} \simeq -1/\eta_e$,   the wavenumber that exits the Hubble radius
at  the end of inflation.

We further define $x = k\eta$ which allows   us to recast Eq.~(\ref{eq:intoq}) as: 
\begin{eqnarray}
\pcq(k) &=& -\f{i}{8} \epsilon_1^4
\left(\ps^0\right)^2\,\int^{x_{\rm e}}_{x_{\rm i}} \f{\d x}{x^2} 
\left( 1 + \f{i}{x} \right)^2 \mathrm{e}^{2ix}
\bigg[c_1 + \f{c_2}{2}\,x^2\bigg] \nn \\
& &+~\text{complex conjugate}\,.
\end{eqnarray}
Here $c_1 =\ln\left(k_{\rm max}/k_{\rm min}\right)$  and $c_2 = (k_{\rm max}^2 - k_{\rm min}^2)/k^2$.   Note that the limits of the time integral in terms of $x$ become
$x_{\rm i} = k\ei$ and $x_{\rm e} = k\ee$, where $\ei \to -\infty$ and 
$\ee \to 0$. But to be consistent with the definitions of $k_{\rm min}$ and 
$k_{\rm max}$, we choose $\ei = -1/k_{\rm min}$ and $\ee = -1/k_{\rm max}$.
This is understandable as the initial  conformal time  corresponds to the epoch 
when the largest scale of interest was well inside the Hubble radius and
the conformal time at the end of inflation is when the smallest scale exits it.
Thus we can express the limits of the integral over $x$ as
$x_{\rm i} =- k/k_{\rm min}$ and $x_{\rm e} =- k/k_{\rm max}$.
Using the values of $k_{\rm min}$ and $k_{\rm max}$, we obtain the estimates 
of $c_1$ and $c_2$ to be
\begin{eqnarray}
c_1 &=& \ln\left(\f{k_{\rm max}}{k_{\rm min}}\right) =
\ln\left(\f{\ei}{\ee}\right) \simeq 70\,, \\
c_2 &=& \f{(k_{\rm max}^2 - k_{\rm min}^2)}{k^2}
\simeq \left(\f{k_{\rm max}}{k}\right)^2\,.
\end{eqnarray}
Using  the wavenumber corresponding to  the pivot scale of CMB 
$k = k_\ast = 5\times 10^{-2}\,\mpcinv$, we get  $c_2 \simeq 10^{42}$. 
  We perform the integral over $x$ and obtain 
\begin{eqnarray} 
\pcq(k) & \simeq & \f{1}{8} \epsilon_1^4 \left(\ps^0\right)^2\,
\bigg[\left(\f{28}{9} - \f{4 \gamma}{3} - \f{4}{3} \ln(-2x_{\rm e})\right)c_1 + 
3\,c_2 \bigg]\,,
\label{eq:powssr}
\end{eqnarray}
where $\gamma \simeq 0.577$ is Euler-Mascheroni constant. Eq.~(\ref{eq:powssr}) gets its main contribution 
 from the upper limit of the integral $x_{\rm e} \ll 1$.
The contribution from the lower limit of $\eta = \ei$ is neglected since the 
integrand is highly oscillatory in that regime and hence will be suppressed 
on integration over $x$.

We have already identified $c_1$ and $c_2$ as contributions from the super-Hubble
and sub-Hubble scales, respectively. 
We can clearly see $c_1$ to be source of infrared divergence from its behaviour as   $x\rightarrow 0$. In our
case, $x_{\rm e}=k/k_{\rm max}$ and $c_1 \propto \ln(-2x_{\rm e})$. $c_2$  signifies ultra-violet divergence as $k_{\rm max} \rightarrow \infty$. This terms dominates for the expected value of $k_{\rm max}$ (the last scale to exit the Hubble radius before the end of inflation). 
However, as this term arises from sub-Hubble physics which is uncertain and subject to renormalization before it can be 
interpreted, it is normally neglected (for similar treatments in the case
of cubic order action, see Refs.~\cite{Sloth:2006az,Dimastrogiovanni:2008af,Adshead:2008gk,Senatore:2009cf,
Kristiano:2021urj,Tasinato:2023ukp}).  

Neglecting the quadratic divergence involving $c_2$, we focus on term containing the 
logarithmic form of $c_1$ that yields
\begin{eqnarray}
\pcq(k) &\simeq& -\f{1}{6} \epsilon_1^4 \left(\ps^0\right)^2\,c_1\,
\ln\left(2\f{k_\ast}{k_{\rm max}}\right) \sim 10^{-28}\,.
\end{eqnarray}
Using $k = k_\ast$, the CMB pivot scale and the value of power spectrum  $\ps^0(k_\ast) \simeq 2 \times 10^{-9}$, $\pcq(k)$ is found to be extremely 
small when compared to the first order result  $\ps^0(k)$, leading to a 
relative correction of $\pcq(k_\ast)/\ps(k_\ast) \simeq 10^{-19}$.
We did not use any specific model of inflation  for obtaining this estimate, 
but we use only one term from the quartic action, which is chosen to maximize the contribution in the SR-USR-SR model  owing to 
sharp transition between the phases. However, other terms are also on the order $\epsilon_1^4$ and second order in 
the first order power spectrum, so we have reasons to expect that the order of magnitude we obtain  gives  generic one 
loop-level contribution to $\ps(k)$  due to quartic action for slow roll inflation.


\subsection{Model of interest : SR-USR-SR}
We now turn to the model of interest, a brief epoch of USR occurring between
two phases of slow roll. We shall use the slow roll over parameters  $\epsilon_i$s modelled in section~\ref{sec:model}
(Eq.~\eqref{eq:epss}) along with the approximations of $a \simeq -1/(H\eta)$
and $V \simeq 3H^2\Mpl^2$.  This allows us to identify the epochs that contribute significantly to the 
time integral in Eq.~(\ref{eq:pc-def}). In the first and the third SR phase, the contribution is negligible as already noted in the previous section. This allows us to restrict the time integral to the USR phase. This gives us:
\begin{eqnarray}
\pcq(k) &=& \f{ik^3}{2\pi^2}\f{\Mpl^2}{H^2}\epsilon_{1_{\rm i}}
\epsilon_2^{\rm II}\,f^2_k(\ee)
\int_{\eta_1}^{\eta_2} \f{\d \eta}{\eta^3} \left( \f{\eta}{\eta_1} \right)^6 
\left( \pm \f{1}{\sqrt{\pi}\Delta \eta} \mathrm{e}^{-\f{(\eta-\eta_{1,2})^2}{\Delta \eta^2}} \right)
\left[ 1 - 2\f{\eta(\eta-\eta_{1,2})}{\Delta \eta^2}\right] \nn \\
& & \times \left[ \f{2\eta(2\eta-\eta_{1,2})}{\Delta \eta^2 - 2\eta(\eta-\eta_{1,2})}\right] 
\left[ f^\ast_k(\eta)\right]^2 \int \d \ln q\, \ps(q,\eta) +~\text{complex conjugate}\,.~~
\label{eq:powcousr}
\end{eqnarray}
As discussed in section~\ref{sec:model}, we consider sharp transition between different phases in this paper.  In this case, $\Delta \eta \ll 1$, i.e. 
$\exp[-(\eta-\eta_{1,2})^2/\Delta\eta^2]/(\sqrt{\pi}\Delta \eta) \to 
\delta^{(1)}(\eta-\eta_{1,2})$. With this scenario in view, it can be shown  that most of the contribution to Eq.~(\ref{eq:powcousr}) arises from the two transition points.   
This simplifies Eq.~(\ref{eq:powcousr}) to
\begin{eqnarray}
\pcq(k) &=& i\,\f{\Mpl^2}{H^2}\f{\epsilon_{1_{\rm i}}
\epsilon_2^{\rm II}}{\Delta\eta^2}\f{k^3}{2\pi^2}\,f^2_k(\ee)
\bigg[\f{[\zeta_k^\ast(\eta_1)]^2}{\eta_1} \int \d \ln q\, \ps(q,\eta_1)\nn\\ 
& &- \left( \f{\eta_2}{\eta_1} \right)^6\f{[f_k^\ast(\eta_2)]^2}{\eta_2} 
\int \d \ln q\, \ps(q,\eta_2) \bigg] + \text{complex conjugate}\,.
\label{eq: expression pc}
\end{eqnarray}
We then utilize the form of the mode functions $f_k(\eta)$ that we have 
solved in the earlier section (Eqs.~\eqref{eq:vk-I},~\eqref{eq:vk-II} 
and~\eqref{eq:vk-III}) and obtain the expression of
$\pcq(k)$ as
\begin{eqnarray}
\pcq(k) &=& \f{i}{4} \left(\f{H^2}{8\pi^2\Mpl^2\epsilon_{1_{\rm f}}}\right)^2 
\f{\epsilon_2^{\rm II}}{k^3\eta_2\Delta\eta^2}\,{\cal F}^2(\alpha_k,\beta_k,\ee) \nn \\
& & \times 
\Bigg\{ \big({\cal F}^\ast(1,0,\eta_1)\big)^2\left( \f{\eta_2}{\eta_1} \right)^7 
\int \d \ln q \left( 1 + q^2\eta_1^2\right) \nn \\
& & - \big( {\cal F}^\ast(\alpha_k, \beta_k, \eta_2) \big)^2
\Bigg[ \int \d \ln q \left( 1 + q^2\eta_2^2\right) 
+ \int \d \ln q\, \bigg[ 2\vert \beta_q \vert^2 (1+q^2\eta_2^2) \nn \\
& & + \bigg(\alpha^\ast_q\beta_q \mathrm{e}^{2iq\eta_2} (q^2\eta_2^2-1+2iq\eta_2)
+ \alpha_q\beta^\ast_q \mathrm{e}^{-2iq\eta_2} (q^2\eta_2^2-1-2iq\eta_2) \bigg)\bigg]\Bigg]
\Bigg\} \nn \\
& & +~\text{complex conjugate}\,,
\label{eq: Pc_an_ecp}
\end{eqnarray}
where the function ${\cal F}(\alpha_k,\beta_k,\eta)$ is defined for convenience as
\begin{equation}
{\cal F}(\alpha_k,\beta_k,\eta) = \alpha_k\, \mathrm{e}^{-ik\eta}(k\eta-i)
+ \beta_k\, \mathrm{e}^{ik\eta}(k\eta+i)\,.
\end{equation}
We can notice a few points about the structure of $\pcq(k)$. 
The main contribution to  $\pcq(k)$ is entirely due to the transitions between the slow roll and USR
regimes captured as a Dirac delta function in $\epsilon_3$. We notice  that 
when we set $\eta_1=\eta_2$, $\alpha_k = 1$ and $\beta_k = 0$, so that the terms 
inside the curly braces cancel each other,  this contribution vanishes leaving 
behind the subdominant behaviour due to the small but non-zero values of 
$\epsilon_i$s as seen in the case of slow roll.
For a finite duration of USR, $\pcq(k) \propto 1/\Delta\eta^2$. 
This indicates that in the limit of a sharp transition with $\Delta \eta\to 0$, 
$\pcq(k)$ becomes very large. 
It is essentially because of the higher order slow roll parameter $\epsilon_4 
\propto 1/\Delta\eta^2$.
On the other hand, when the transition is smoothed using a finite value 
of $\Delta \eta$, it attenuates the amplitude of $\pcq(k)$ and in the 
extreme limit of $\Delta\eta \to \infty$, $\pcq(k) \to 0$  
(see e.g. \cite{Franciolini:2023lgy} who have  recently studied the 
impact of the smoothness of this  transition on the power spectrum 
correction for cubic order action).

The structure of the  integral over the loop momentum $q$, evaluated
at $\eta_1$ or $\eta_2$, follows the logic of the discussion  in the slow roll case. 
We find that there arise two kinds of divergences in the limits 
of integration. In the upper limit, the divergence is dominated by the term 
$q^2\eta^2$ as $q \to \infty$ (the ultra-violet divergence whose origin is the shape of mode function  on sub-Hubble scales), whereas in the lower limit the divergence is of 
the form $\ln[q_{\rm min}\eta]$ as $q_{\rm min} \to 0$ (the infra-red divergence that arises from physics on super-Hubble scales).These divergences have been studied in the literature, 
mainly in the context of slow roll inflation~\cite{Maldacena:2002vr,Collins:2005nu,
Gerstenlauer:2011ti,Seery:2010kh}. The contribution 
owing  to sub-Hubble  modes at any given 
time has to be regularized and  renormalized and can not be treated as observable.
Therefore, we restrict the range of integral over loop momentum to 
the super-Hubble  modes at $\eta_1$ and $\eta_2$, i.e. $\vert q\,\eta_{1,2} \vert <1$
for the respective integrals.
We relegate further discussion regarding the nature of the quadratic divergence 
to Appendix~\ref{app:divergence}.

Thus, we arrive at $\pcq(k)$, with integrals over $q$ at $\eta_1$ and $\eta_2$ 
performed only with contribution from super-Hubble modes to be
\bea
\pcq(k) &\simeq & 
\f{i}{4}\left(\f{H^2}{8\pi^2\Mpl^2\epsilon_{1_{\rm f}}}\right)^2 
\f{\epsilon_2^{\rm II}}{k^3\eta_2\Delta\eta^2}\,
{\cal F}^2(\alpha_k,\beta_k,\ee) \nn \\
& & \times\Bigg\{ \big({\cal F}^\ast(1,0,\eta_1)\big)^2\left( \f{k_1}{k_2} \right)^7 
\bigg[ \ln \left( \f{k_1}{k_{\rm min}} \right) 
+ \f{1}{2}\left( 1- \f{k_{\rm min}^2}{k_1^2}\right) \bigg] \nn \\
& & - \big( {\cal F}^\ast(\alpha_k, \beta_k, \eta_2) \big)^2
\Bigg[ \left(\f{k_1}{k_2}\right)^6 \bigg[\ln \left( \f{k_2}{k_{\rm min}}\right) - \f{1}{10}\left( 1- \f{k_{\rm min}^2}{k_2^2}\right)\bigg] \nn \\
& &-\bigg[\f{2}{5}\left(\f{k_1}{k_2}\right) -\left(\f{k_1}{k_2}\right)^4\bigg]
\bigg[1-\left(\f{k_{\rm min}}{k_2}\right)^2\bigg]
\Bigg]\Bigg\} +~\text{complex conjugate}\,, \nn \\
\label{eq:Pc_superH}
\eea
where we have set $k_1=-1/\eta_1$ and $k_2=-1/\eta_2$.

We shall explore the behaviour of this $\pcq(k)$ in terms of the three 
parameters that characterize the model, namely, the onset of USR, now 
parameterized using the corresponding wavenumber $k_1$, the duration of USR
which we shall measure in e-folds as $\Delta N$ and the smoothness of the 
transition between slow roll and USR epochs, $\Delta \eta$.
The parameters $k_2$ and $k_1$ are related as $k_2 = k_1\,\exp(\Delta N)$.


\section{Results}\label{sec:results}

\begin{figure}
\centering
\includegraphics[width=14.5cm]{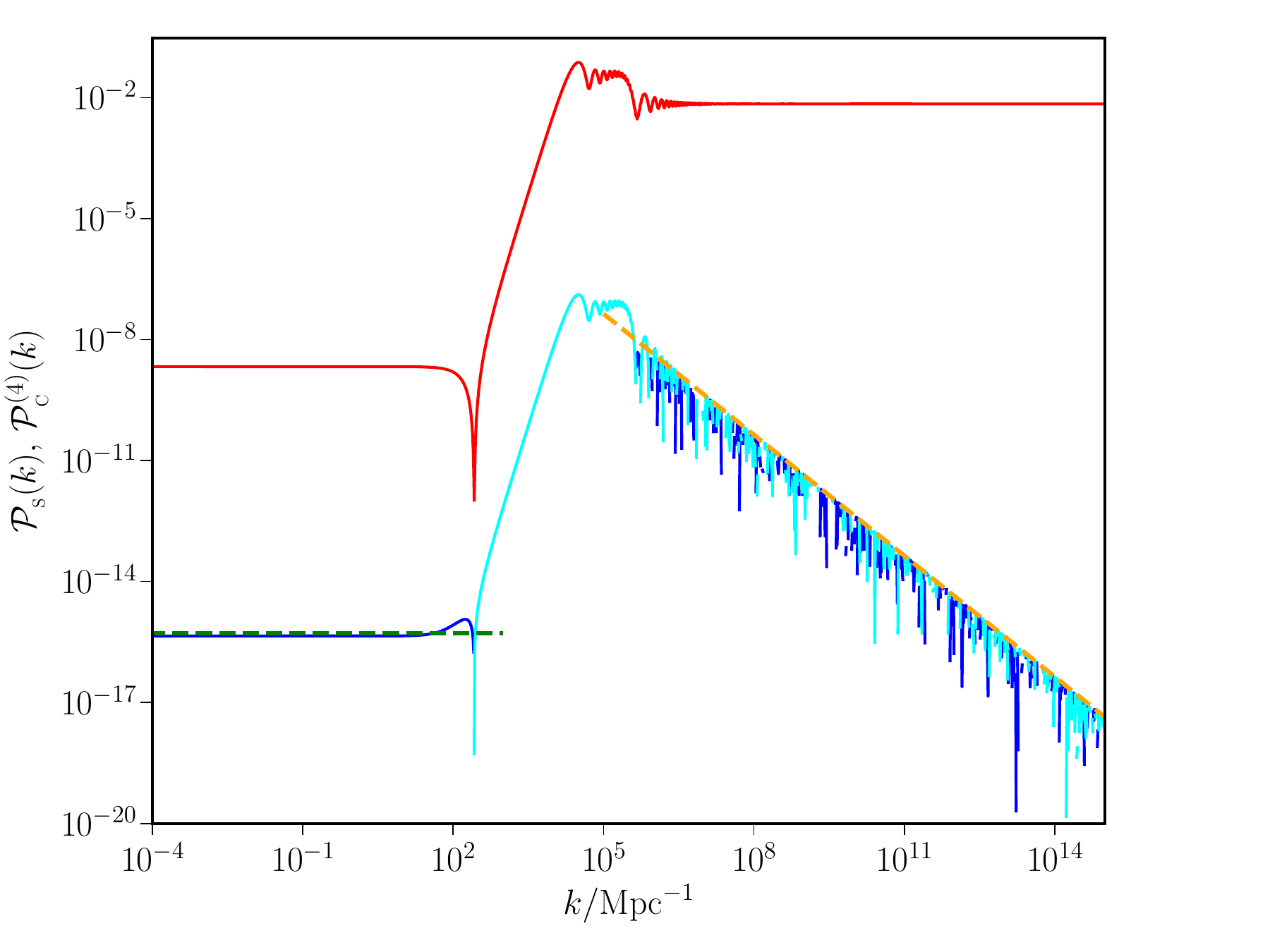}
\caption{The first order power spectrum $\ps(k)$  (in red) and  the loop correction  $\pcq(k)$ (in blue) are displayed 
for the following parameters: $H=1.3 \times 10^{-5}\,\Mpl, \epsilon_{1_{\rm i}}=10^{-3},\,\epsilon_2^{\rm II}=-6,\, \Delta\eta = 10^{-3}\,{\rm Mpc},\,k_1=10^4\,\mpcinv,\, \Delta N = 2.5,\,
k_{\rm min}={10^{-6}}\,\mpcinv,\, k_{\rm max}=10^{20}\,\mpcinv$. The negative values of  $\pcq(k)$ are shown in cyan. 
We also plot the asymptotic expressions obtained in Eqs.~\eqref{eq:pc-large-k}  (dashed green)
and the envelope of Eq.~\eqref{eq:asymlargek} (dashed orange)  to compare then with the exact expressions
in respective regimes. We note that the oscillatory features seen at small scales in the loop correction  $\pcq(k)$ are also captured well by Eq.~\eqref{eq:asymlargek} }\label{fig:ps-pc-rep}
\end{figure}

We begin by examining the general shape of the one-loop correction $\pcq(k)$ and understanding the
prominent features therein. We present the behaviour of  $\pcq(k)$ against the first-order power spectrum $\ps(k)$ (Eq.~(\ref{eq:powspec1}))
in Fig.~\ref{fig:ps-pc-rep}. We have made suitable choice of parameter values such 
that we obtain correct normalization over CMB scales and sufficient enhancement
over small scales. The first-order power spectrum  $\ps(k)$ over large scales with $k \ll k_1$ 
and small scales of $k \gg k_2$ agree with the asymptotic expressions mentioned 
in Eqs.~\eqref{eq:ps-large-k} and~\eqref{eq:ps-small-k}, respectively.

We attempt to obtain similar expressions characterizing the asymptotic behaviour
of the correction  $\pcq(k)$ over large and small scales. We first consider $\pcq(k)$ over large 
scales of $k \ll k_1 < k_2$. In such a limit, we obtain 
\bea
\pcq(k \ll k_1) &\simeq & -\f{\epsilon_2^{\rm II}}{3\,k_1^2\Delta\eta^2}
\left(\f{H^2}{8\pi^2\Mpl^2\epsilon_{1_{\rm i}}}\right)^2
\Bigg\{ \left[ 2\l(\f{k_2}{k_1}\r)^3 -1 \right] \l[ \ln\l(\f{k_1}{k_{\rm min}}\r) + \f{1}{2} \r] \nn \\
& &-\l(\f{k_2}{k_1}\r)^4\l[\ln\l(\f{2k_2}{k_1}\r)+\gamma-\f{1}{6}\r]\Bigg\}\,, \nn \\
& \simeq &-\f{2\,\epsilon_2^{\rm II}}{3\,k_1^2\Delta\eta^2}
\left(\f{H^2}{8\pi^2\Mpl^2\epsilon_{1_{\rm i}}}\right)^2
\l(\f{k_2}{k_1}\r)^3 \ln\l(\f{k_1}{k_{\rm min}}\r)\,.
\label{eq:pc-large-k}
\eea
Eq.~(\ref{eq:pc-large-k}) shows  that $\pcq(k)$ is scale invariant and 
positive over large scales,
when $k_1/k_{\rm min}$ is  much  larger as  compared to $k_2/k_1$. It is seen in Figure~\ref{fig:ps-pc-rep}   
that  Eq.~\eqref{eq:pc-large-k}  captures well the asymptotic form of Eq.~(\ref{eq:Pc_superH}) in the large scale limit.
Using
$\ps(k \ll k_1) \simeq H^2/(8\pi^2\Mpl^2\epsilon_{1_{\rm i}})$,  the ratio of $\pcq(k)$ to $\ps(k)$ over these scales is
\begin{equation}
\f{\pcq(k)}{\ps(k)} = 
-\f{2\,\epsilon_2^{\rm II}}{3\,k_1^2\Delta\eta^2}
\l(\f{H^2}{8\pi^2\Mpl^2\epsilon_{1_{\rm i}}}\r)
\l(\f{k_2}{k_1}\r)^3 \ln\l(\f{k_1}{k_{\rm min}}\r)\,.
\end{equation}
For the values of parameters we have used in  Fig.~\ref{fig:ps-pc-rep}, this ratio is much smaller than
unity. 
In this case,  the loop-level contribution $\pcq(k)$ poses no problem
to the validity of the perturbation theory. 
We study other cases of interest in the next sub-sections.  

The shape of $\pcq(k)$ at  $k \simeq k_1$ mimicks the enhancement of $\ps(k)$ 
with the slope of $k^4$. An interesting feature to note is that $\pcq(k)$ 
contains a  null at the same scale $k = k_{\rm dip}$ as the first-order power spectrum $\ps(k)$ (for details see e.g. \cite{Ozsoy:2021pws,Balaji:2022zur} and references therein).
This suggests that the null in $\ps(k)$ shall not get altered due to  the addition of 
$\pcq(k)$. Another feature of $\pcq(k)$ is that it is not guaranteed to be 
positive over all scales. In fact we see it crossing zero at $k_{\rm dip}$ and then
becoming negative over the range where it is enhanced in amplitude. Further, we 
observe it altering between positive and negative values over scales of $k \geq k_1$
and hence containing multiple other nulls whenever it crosses zero.
The sign and the oscillatory nature of $\pcq(k)$ arise essentially from the 
relative phase difference when  the mode functions at   two different times,  $\ee$ and  $\eta$, are contracted.
To better understand the origin of the sign  and the overall structure of $\pcq(k)$, we 
rederive the quantity in a slightly different manner in Appendix~\ref{app:pc-struc}\,.

We  next examine the asymptotic form of the correction  $\pcq(k)$ over small scales 
 $k \gg k_2 > k_1$. In this limit, we obtain 
\bea
\pcq(k \gg k_2) & \simeq & -\f{\,\epsilon_2^{\rm II}}{2kk_2\Delta\eta^2} 
\left(\f{H^2}{8\pi^2\Mpl^2\epsilon_{1_{\rm f}}}\right)^2
\Bigg\{\l[\ln\l(\f{2k_2}{k_1}\r) + \gamma - \f{1}{6}\r]\sin\l(\f{2k}{k_2}\r) \nn \\
& & -\l(\f{k_1}{k_2}\r)^5
\l[\ln\l(\f{k_1}{k_{\rm min}}\r) + \f{1}{2}\r]\sin\l(\f{2k}{k_1}\r) \Bigg\},
\label{eq:asymlargek}
\eea
where $\gamma \simeq 0.577$ is the Euler-Mascheroni constant. We can see that the 
first term within the curly braces dominates the second term, due to the factor of
$(k_1/k_2)^5$\,. The overall prefactor that forms the envelope of oscillations 
suggests that $\pcq(k) \propto 1/k$ over small scales and it matches the complete
$\pcq(k)$ as can be seen in Fig.~\ref{fig:ps-pc-rep}.
The ratio of $\pcq(k)$ to $\ps(k)$ in this regime is
\begin{equation}
\f{\pcq(k)}{\ps(k)} \simeq -\f{\epsilon_2^{\rm II}}{2\,k\,k_2\,\Delta\eta^2}
\left(\f{H^2}{8\pi^2\Mpl^2\epsilon_{1_{\rm f}}}\right)\,
\l[ \ln\l(\f{2\,k_2}{k_1}\r) + \gamma -\f{1}{6} \r]\,
\label{eq:pc-small-k}
\end{equation}
where we have used $\ps(k \gg k_2) = H^2/(8\pi^2\Mpl^2\epsilon_{1_{\rm f}})$.
We also notice that  $\pcq(k)$ scales as $1/k_2^2$
and $1/k_2$ in Eqs.~\eqref{eq:pc-large-k}
and~\eqref{eq:pc-small-k}, respectively. 
These terms suggest that $\pcq(k)$, though sub-dominant for large $k_2$, the case shown
in Fig.~\ref{fig:ps-pc-rep}, can become large and comparable to $\ps(k)$ for smaller values 
of $k_2$.

An important feature of Eq.~(\ref{eq:asymlargek}) is that the correction $\pcq(k)$ can become negative at small scales.  In Appendix ~\ref{app:pc-struc},  we investigate the structure of the correction $\pcq(k)$  in detail with particular focus on the origin of its sign.  The power spectrum,
the sum of first-order power spectrum and the correction, is positive definite so the correction can be negative if it doesn't 
dominate the first-order contribution. However, we later find cases in which the correction can be negative and dominate the first-order power spectrum. This signals
the breakdown of both perturbative expansion and the positivity of the power spectrum and indicates that we are not accounting for 
all the relevant terms in the perturbative expansion. A detailed investigation that takes into account the impact of all the terms
whose inclusion is needed to justify the perturbative expansion at small scales is left to a future work. 

In next subsections, we  further study the dependence of  $\pcq(k)$  on the three defining 
parameters of the model: (a) the onset of the USR phase,  $\eta_1 = -1/k_1$, (b) the duration of the USR phase, $\Delta N$\, and (c) the smoothness of transition,  $\Delta \eta$.
We carry out our analysis for  three separate cases that have been motivated by different contexts in the literature---(i) late onset of USR that affects 
the power spectrum over small scales and is relevant for the production of   primordial black holes, (ii) an intermediate  regime in which the power enhancement is constrained by CMB spectral distortion, (iii) the  early onset of USR which affects the 
spectrum over large scales and has been studied for understanding the CMB angular power spectra at large angular scales.

\subsection{Late onset of USR}

We first consider the scenario where the USR epoch occurs during later stages of inflation
as is required for enhancement of power over small scales to induce significant 
production of PBHs~(see for recent reviews,~\cite{Ragavendra:2023ret,Ozsoy:2023ryl} 
and references therein). The parameters such as $H$ and $\epsilon_{1_{\rm i}}$ are
chosen such that the large scales exiting the Hubble radius  in the initial 
slow roll epoch have their $\ps(k)$ normalized to about $2\times 10^{-9}$ around
the pivot scale of CMB $k_\ast = 5 \times 10^{-2}\,\mpcinv$. This initial slow roll
epoch also ensures that  the power spectrum $\ps(k)$ is nearly scale invariant at large scales which is 
required by CMB anisotropy data \cite{Planck:2019kim}.
The onset of USR is set at a sufficiently later stage so that  both the amplitude and shape of the  power spectrum at  such scales
are   unaffected. The duration of USR  that follows 
is chosen suitably such that the amplitude of the  spectrum is enhanced to about 
$10^{-2}$ at its peak. The last slow roll epoch that follows leads to yet another 
scale invariant part of $\ps(k)$ over small scales.

In such a case, we compute $\pcq(k)$ (Eq.~\eqref{eq:Pc_superH})
with $\eta_1$ chosen to be sufficiently late such that $k_1 \geq 10^4\,\mpcinv$\,. 
We choose the values of other relevant parameters to be $H=10^{-5}\,\Mpl$,
$\epsilon_{1_{\rm i}}=7\times10^{-4}$ to achieve appropriate normalization of $\ps(k)$
over CMB scales.
We know that $k_2 = k_1\exp(\Delta N_{\rm USR})$, and we set 
$\Delta N_{\rm USR}=2.5$ so that the amplitude of the peak in the spectrum is
about $\ps(k) \sim 5\times 10^{-2}$, as is required to produce significant 
population of PBHs. 
We also set the value of $\Delta \eta=10^{-3}\,{\rm Mpc}$ to mimic instantaneous
transition and later vary this parameter to determine the impact  of smoothening the transition. 

\begin{figure}
\centering
\includegraphics[width=14.5cm]{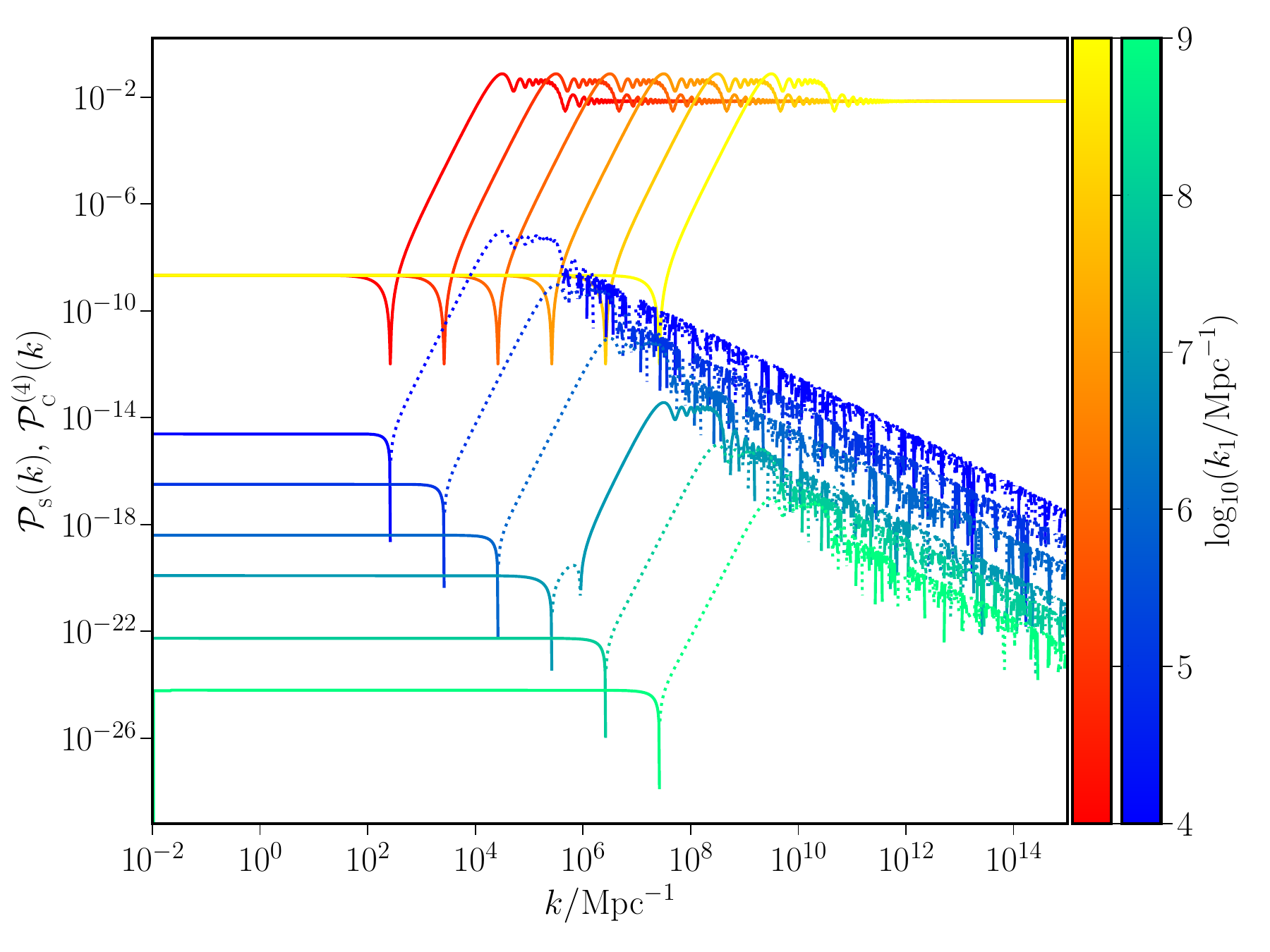}
\caption{The quantities $\ps(k)$ (in shades of red to yellow) and $\pcq(k)$ 
(in shades of blue to green) are illustrated here for the case of late onset of USR.
Note that $\pcq(k)$ contains both positive and negative values (plotted in solid
and dotted lines).
The onset of USR $\eta_1$ is varied over a range such that the corresponding 
wavenumber $k_1=-1/\eta_1$ takes the values, $k_1=10^4,\,10^5,\,10^6,\,10^7,\,10^8$ 
and $10^9\,\mpcinv$. The duration of USR is set to be $\Delta N= 2.5$, so that 
the enhancement in $\ps(k)$ over small scales is sufficiently large to produce 
PBHs, while ensuring appropriate CMB normalization over large scales. 
We observe that $\pcq(k) \propto 1/k_1^2$ and hence earlier the onset, larger is 
the correction. However, the correction is highly sub-dominant even for the case 
when the enhancement of power occurs as early as with $k_1=10^4\,\mpcinv$.}
\label{fig:pc-k1-late}
\end{figure}

In Figure~\ref{fig:pc-k1-late}, we present $\pcq(k)$, varied over a range of $k_1$,
that correspond to  different onsets of USR. In this case,  the amplitude of
$\pcq(k)$  over the entire range of $k$ is  seen to be much smaller than $\ps(k)$. This has a 
crucial implication for the scenario of production of PBHs. The loop-level
contribution arising from quartic action to the first-order power spectrum spectrum is sub-dominant
and does not alter  the prediction for observables. As  $\pcq(k) \propto 1/k_2^2 \propto 1/k_1^2$,  the correction  becomes even less significant  when the onset of USR is pushed 
to  later stages of inflation.

\begin{figure}
\centering
\includegraphics[width=14.5cm]{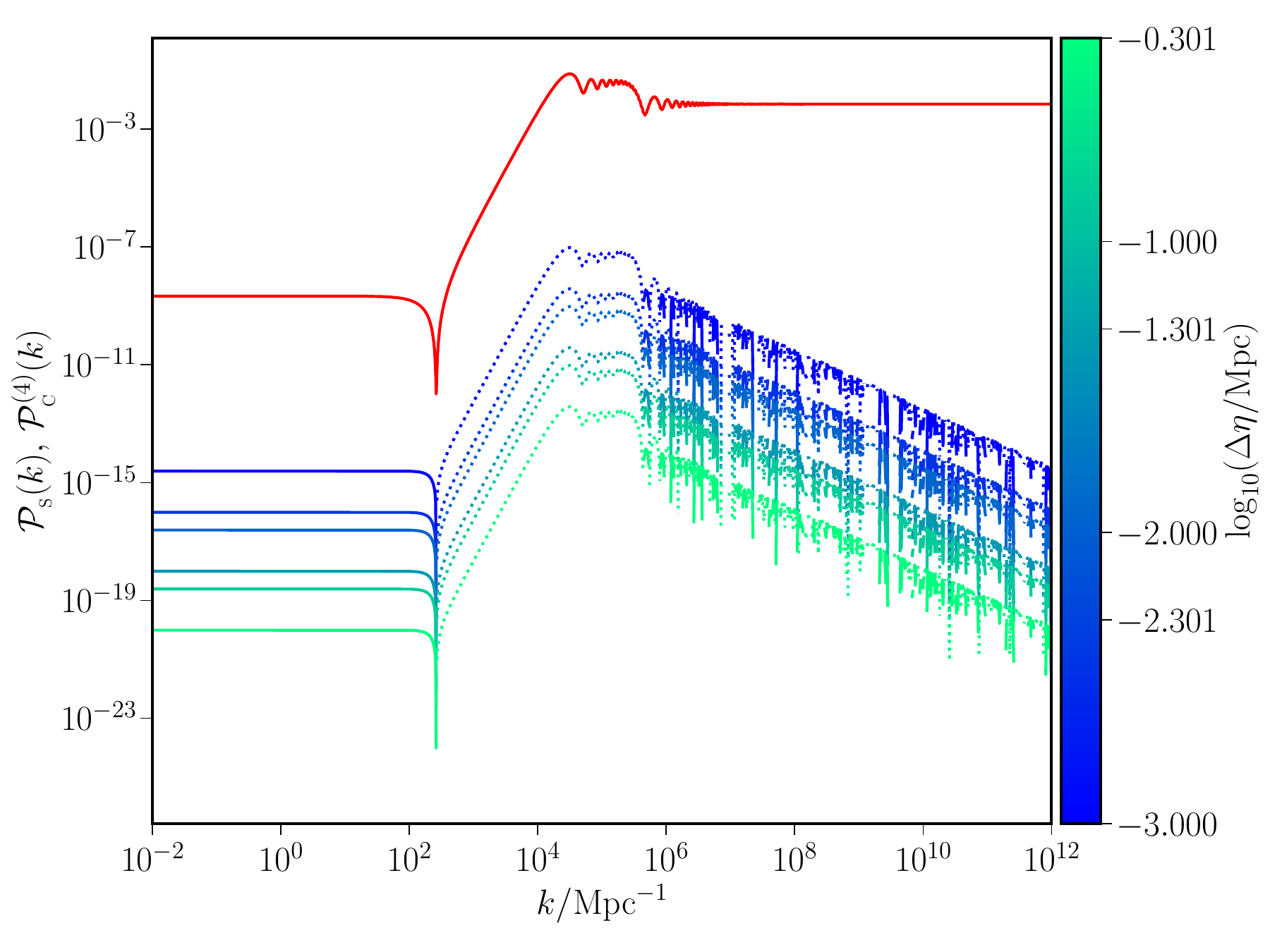}
\caption{The $\pcq(k)$ (in shades of blue to green) is presented against $\ps(k)$
(in red) across the variation in smoothness parameter $\Delta \eta$. 
The parameter is varied over the values 
$\Delta\eta= 0.5,~0.1,~5 \times 10^{-2},~10^{-2},~5\times 10^{-3},$ and 
$10^{-3}\,{\rm Mpc}$. The onset of USR is set such that $k_1=10^{4}\,\mpcinv$ and 
the duration is taken to be $\Delta N=2.5$ to obtain this plot.
We observe that the $\pcq(k)$ falls as $1/\Delta \eta^2$ as expected and contains
both positive and negative values as seen before (in solid and dotted lines).}
\label{fig:pc-deta-late}
\end{figure}

In Figure~\ref{fig:pc-deta-late}, we present the behavior of $\pcq(k)$ with respect
to the smoothening parameter $\Delta \eta$. We observe that the overall amplitude of 
$\pcq(k)$ is suppressed with larger values of $\Delta \eta$. This is to be expected
from Eq.~\eqref{eq:Pc_superH}, where we can clearly see $\pcq(k) \propto 1/\Delta\eta^2$.
We should note that the smoothening also mildly affects the shape of the spectra as
it suppresses the oscillations present in both $\ps(k)$ and $\pcq(k)$. 
More importantly, modelling smoothness in alternative ways, such as in terms of e-folds,
may lead to significant changes in the behavior of $\pcq(k)$. We discuss this aspect
in detail in App.~\ref{app:smooth}.

\subsection{Intermediate onset of USR}
We  consider a scenario where the phase of USR occurs just after the large
scales probed by Planck ($k \simeq 10^{-4}\hbox{--} 0.2 \, \rm Mpc^{-1}$) exit the Hubble radius. In this case, the parameters are chosen 
such that the modes with $k \ll k_1$ have scalar power normalized to
$\ps^0 \simeq 2.1 \times 10^{-9}$ and the enhancement occurs over scales of
$1\,\mpcinv < k < 10^3\,\mpcinv$\,. We restrict the duration of USR so that
$\ps(k)$ at its maximum is well within the bound on its amplitude $\ps(k) < 10^{-5}$, 
as imposed by FIRAS due to CMB spectral distortions (e.g. \cite{Chluba:2015bqa} and references therein).

\begin{figure}
\centering
\includegraphics[width=14.5cm]{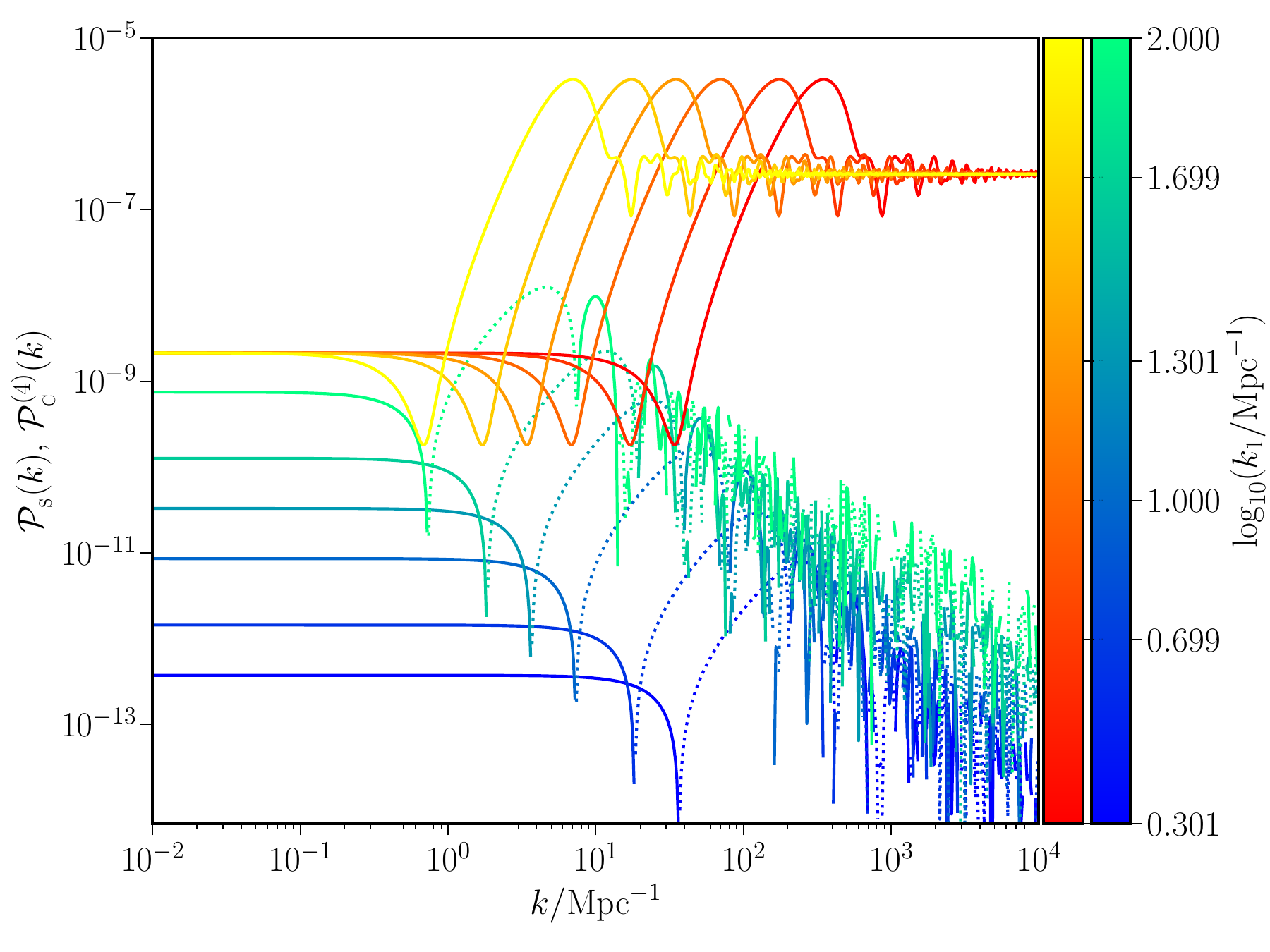}
\caption{We present $\ps(k)$ (in shades of red to yellow) and $\pcq(k)$ (in shades
of blue to green) for the case of onset of USR just after the exit of CMB scales 
(intermediate onset of the USR phase). 
The epochs of the onset of  USR phase ($\eta_1 = 1/k_1$) correspond  $k_1= 2,\,5,\,10,\,20,\,50,$ and $10^{2}\,\mpcinv$.
The values of $H$ and $\epsilon_{1_{\rm i}}$ are set to values as given in 
Fig.~\ref{fig:ps-pc-rep} to ensure correct  normalization over CMB scales.
The duration of USR is set as $\Delta N= 0.8$ so that the enhancement of scalar 
power is within the bound of $\ps(k) < 10^{-5}$ arising from spectral distortions 
constraint as imposed by FIRAS~(e.g. \cite{Chluba:2015bqa}).}
\label{fig:pc-k1-interm}
\end{figure}

We present the behaviour of $\pcq(k)$ against $\ps(k)$  in 
Fig.~\ref{fig:pc-k1-interm}. The parameter $k_1$  is varied to illustrate its 
effect on the first-order power spectrum and the one-loop correction. 

In this case,  the amplitude of $\pcq(k)$ can become comparable to the first-order power spectrum. For $k_1 = 2\,\mpcinv$, the correction 
is nearly  $30\%$ of $\ps(k)$ over the CMB scales. This 
effect is essentially due to the dependence of $\pcq(k) \propto 1/k_1^2$ (Eq.~(\ref{eq:pc-large-k})). However, as Eq.~(\ref{eq:pc-large-k}) shows,  both the correction  $\pcq(k)$ and the first-order power spectrum  $\ps(k)$
are scale invariant for  $k < k_1$.  As  this  correction only alters the overall amplitude of the power spectrum,  its addition is 
degenerate with the normalization of $\ps(k)$. 

The behaviour of the correction at smaller scales is more interesting. 
At the location of dip in the spectrum, $\pcq(k)$ crosses zero and it becomes 
negative over the range of enhancement of $\ps(k)$. This behaviour could alter 
the shape of the complete spectrum around the dip if occurred over intermediate
scales and needs closer attention~\cite{Ozsoy:2021pws,Balaji:2022zur}.

Further, if this scenario of intermediate onset of USR is realized  through a
realistic potential,  there could  be non-trivial effects of $\pcq(k)$ on 
the running of the spectral index, even
over larger scales such as $k_\ast \geq \times 10^{-2}\,\mpcinv$. 
These effects can, in principle, be probed and constrained using the current 
CMB data and we can gain insights about the onset and duration of USR epoch in this
case.

\subsection{Early onset of USR}

We turn to the scenario where the epoch of USR occurs during early stages of 
inflation. This scenario has been considered in literature in the context of
suppressing scalar power over large scales and thereby improving the fit to the
data of CMB anisotropies over lower 
multipoles~\cite{Jain:2008dw,Qureshi:2016pjy,Ragavendra:2020old}.
In this case, the initial slow roll phase occurs before the 
observable scales leave the Hubble radius. The USR phase commences
when the largest scales of interest, such as $k \sim 10^{-4}\,\mpcinv$, leave the 
Hubble radius and  ends  when the scales 
$k \geq k_\ast$ leave the Hubble radius.
If the parameters such as $H$ and $\epsilon_{1_{\rm f}}$ are chosen suitably,
then $\ps(k)$ around $k_\ast$ is appropriately normalized. Such an USR epoch 
effects a suppression over scales larger than $k_\ast$\,, which could results in better fitting  of the CMB data 
at  low multipoles.

We compute $\pcq(k)$   with large values 
of $\eta_1$ such that $k_1 \leq 10^{-2}\,\mpcinv$\,. The other parameters are chosen   to be: $H=10^{-8}\,\Mpl$,
$\epsilon_{1_{\rm i}}=10^{-3}$,  to achieve the required normalization of $\ps(k)$ 
at $k_\ast$. We first set  $\Delta N_{\rm USR}=2.5$ and alter this value to  determine the extent of 
fall of power over large scales. As before, $k_{\rm min} = 10^{-6}\,\mpcinv$ and  $\Delta \eta=10^{-3}\,{\rm Mpc}$ to mimic instantaneous transition;  $\Delta \eta$ is later varied   to discern the impact  of smoothening on $\pcq(k)$.

\begin{figure}
\centering
\includegraphics[width=14.5cm]{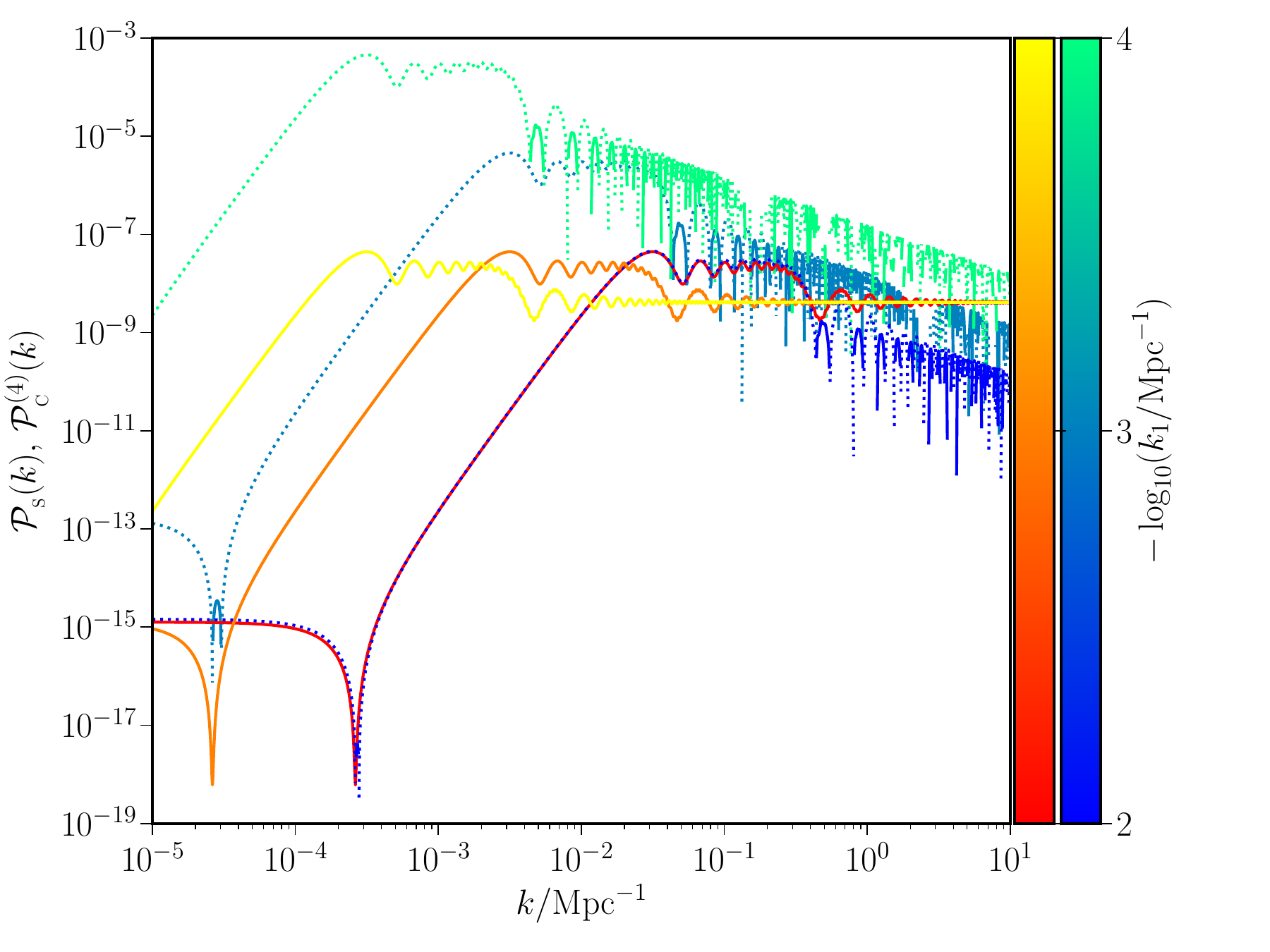}
\caption{The quantities $\ps(k)$ (in shades of red to yellow) and $\pcq(k)$ 
(in shades of blue to green) are presented for the case of early onset of USR.
The onset of USR $\eta_1$ is varied over a range so that $k_1=-1/\eta_1$ takes 
the values, $k_1=10^{-2},~10^{-3}$ and $10^{-4}\,\mpcinv$. 
The duration of USR is set as $\Delta N=2.5$ and $\Delta \eta = 10^{-3}\,$ Mpc.
Other parameters have been chosen such that the spectrum is normalized to about 
$2 \times 10^{-9}$ at $k_\ast$.
Due to the behaviour of $\pcq(k) \propto 1/k_1^2$, we see that $\pcq(k)$ can
become comparable to or even dominant over $\ps(k)$ for sufficiently small $k_1$.}
\label{fig:pc-k1-early}
\end{figure}

We illustrate the effect of variation of $k_1$ on $\pcq(k)$ in Fig.~\ref{fig:pc-k1-early}.
The shape of $\pcq(k)$ is along the expected lines. However, 
its amplitude is comparable to $\ps(k)$ for $k_1 = 10^{-2}\,\mpcinv$ and grows 
to become dominant over $\ps(k)$ as $k_1$ decreases to $10^{-4}\,\mpcinv$. 
As noted above, this is owing to the dependence of 
$\pcq(k) \propto 1/k_2^2 \propto 1/k_1^2$. 

The correction $\pcq(k)$ dominates the first-order term in such cases 
and  suggests a breakdown of the perturbation theory.
Moreover, the sign of the total power spectrum $\ps(k)+\pcq(k)$ can turn 
negative as seen in Figure~\ref{fig:pc-k1-early}. As discussed above, this implies that there ought to be contributions 
from higher-order loops to restore  positivity of the total power spectrum  spectrum at 
all scales. 

\begin{figure}
\centering
\includegraphics[width=14.5cm]{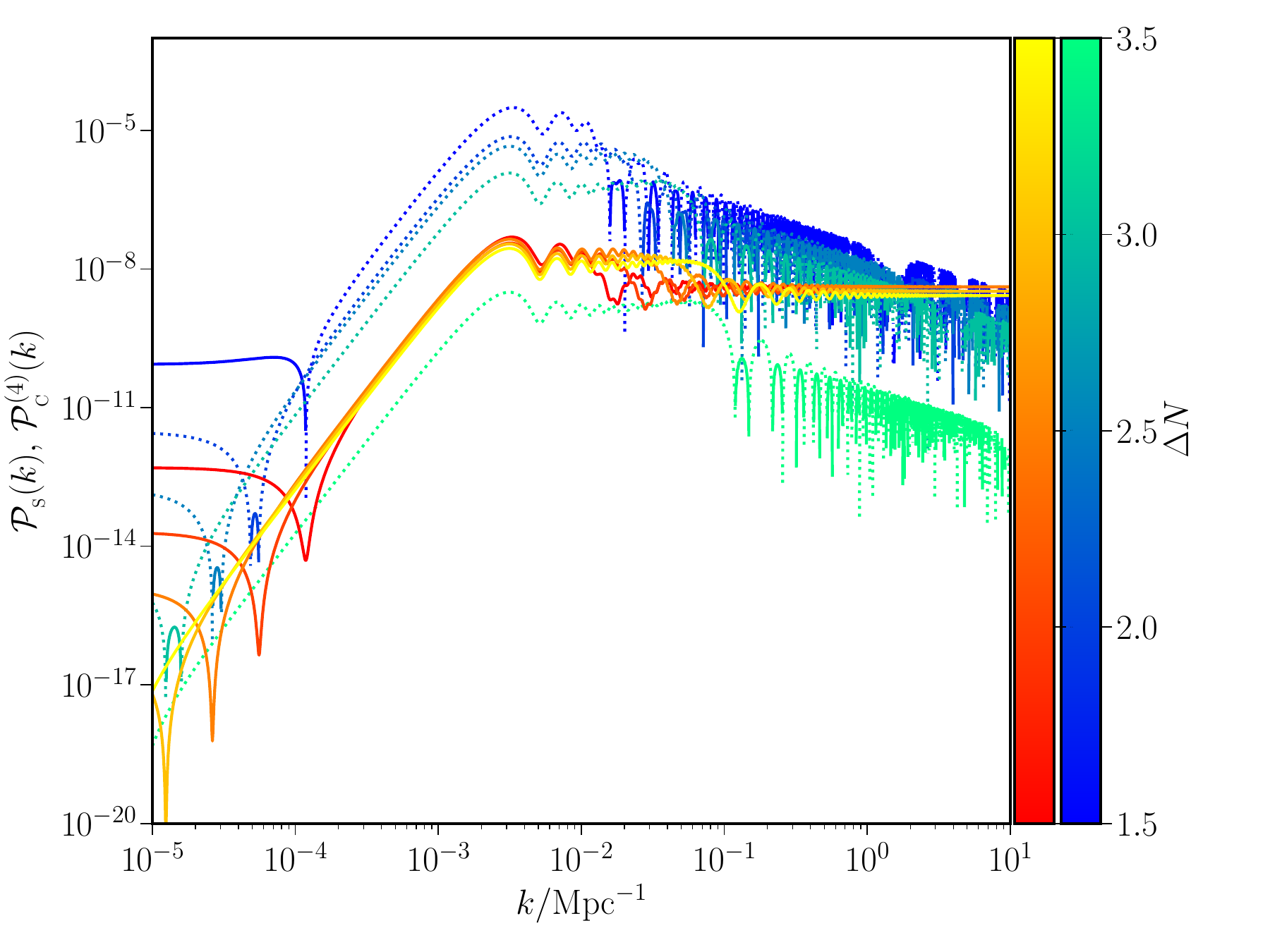}
\caption{The quantities $\ps(k)$ (in shades of red to yellow) and $\pcq(k)$ 
(in shades of blue to green) are plotted here for the case of early onset of USR.
The duration of USR is varied as $\Delta N=1.5,\,2,\,2.5,\,3$ and $3.5$ e-folds.
The onset of USR is set such that $k_1=10^{-3}\,\mpcinv$ and 
$\Delta \eta = 10^{-3}\,$ Mpc to obtain this plot.}
\label{fig:pc-dN-early}
\end{figure}

We examine the behaviour of $\pcq(k)$ with respect to $\Delta N$
in Fig.~\ref{fig:pc-dN-early}. Unlike the previous case, where $\Delta N$
was fixed to enhance power sufficiently to produce PBHs, in this case it
can be varied to explore different levels of suppression over large scales.
We vary $\Delta N$ over a range of 1.5 to 3.5 along with $H$\footnote{
$H$ is varied along with variation of $\Delta N$ as $H \propto \exp(-3\Delta N)$ 
to retain the ratio of $H^2/(8\pi^2\epsilon_{1_{\rm f}})$ as a constant in this 
case.} so that we retain the normalization of $\ps(k)$ around $k_\ast$. 
We observe that both $\ps(k)$ and $\pcq(k)$ suffer stronger suppression over 
large scales with larger values of $\Delta N$.
However, an interesting effect we observe is that the overall amplitude of 
$\pcq(k)$ is enhanced with a decrease in $\Delta N$. This can be understood
from Eq.~\eqref{eq:pc-small-k} as $\pcq(k) \propto (1/k_2)\ln(k_2/k_1)$ for 
$k \gg k_2$\,.
For a fixed $k_1$, it means that $\pcq(k) \propto \Delta N\exp(-\Delta N)$\,. 
This suggests that shorter is the duration of USR, more pronounced is the 
effect of $\pcq(k)$ over these scales.


\begin{figure}
\centering
\includegraphics[width=14.5cm]{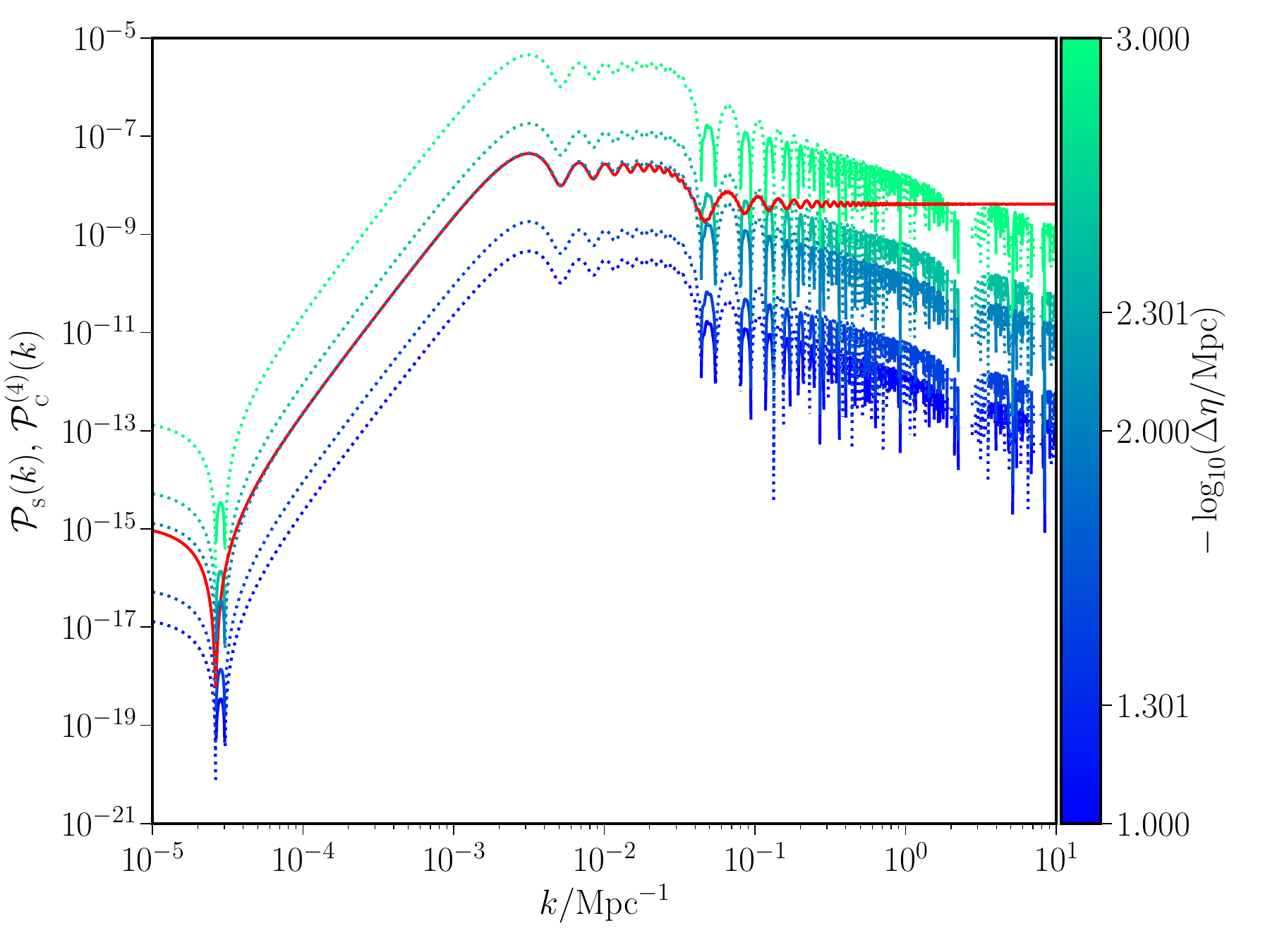}
\caption{The quantities $\ps(k)$ (in red) and $\pcq(k)$ 
(in shades of blue to green) are plotted here for the case of early onset of USR.
The smoothness parameter is varied as 
$\Delta\eta= 0.5,~0.1,~5\times 10^{-2},~10^{-2},~5\times 10^{-3}$ and 
$10^{-3}$\,Mpc.
The onset of USR is set such that $k_1=10^{-3}\,\mpcinv$ and $\Delta N= 2.5$
in obtaining this plot.}
\label{fig:pc-deta-early}
\end{figure}

Lastly, we present the effect of smoothening on the amplitude of $\pcq(k)$ in
Fig.~\ref{fig:pc-deta-early}. As we observed in the previous case, the amplitude
of $\pcq(k)$ is suppressed as $1/\Delta \eta^2$ and for a given $k_1$ and
$\Delta N$, $\pcq(k)$ can be made sub-dominant to $\ps(k)$ by increasing the
smoothening of the transition.

\section{Discussion and outlook}\label{sec:conc}

We have investigated the loop-level contribution to scalar power spectrum arising from the action governing the scalar perturbation at the quartic order for 
inflationary models with a brief epoch of USR.
We parameterize the model using three parameters---$k_1$, $\Delta N$ and 
$\Delta \eta$\, that characterize 
 the onset  of  the USR phase, the  duration of USR, and the smoothness of transitions between the  SR 
and USR phases, respectively. We consider three scenarios corresponding to  early, intermediate, and late transition to the USR phase. We find  the loop correction to the power spectrum  $\pcq(k)$ is  sub-dominant  to the first-order contribution 
if the  USR phase  occurs at  later stages of the  inflationary epoch (Figures~\ref{fig:pc-k1-late} and~\ref{fig:pc-deta-late}). 
When  USR occurs at intermediate and  early stages, we find that the correction 
 can become comparable to or dominate the first-order contribution, which can leave interesting  signatures on cosmological observables but it also  could pose  a challenge to the validity of  perturbative expansion (Figures~\ref{fig:pc-k1-interm}--\ref{fig:pc-deta-early}).

We show that $\pcq \propto 1/k_1^2$ for $k \ll k_1$ (Eq.~(\ref{eq:pc-large-k})) and $\pcq \propto 1/k$ for $k \gg 
k_1$ (Eq.~(\ref{eq:asymlargek})). To ensure that the loop correction remains smaller than the first-order power spectrum, we require the onset of USR  ($\eta_1 = 1/k_1$) such that 
$k_1 < 10^{-2}\,\mpcinv$.
The duration of USR $\Delta N$, plays a more complex role in determining the shape and amplitude of  $\pcq(k)$. It follows from Eq.~\eqref{eq:pc-large-k} that $\pcq(k) \propto \exp(3\,\Delta N)$ 
over large scales whereas $\pcq(k) \propto \Delta N\exp[-\Delta N]$ over small scales (Eq.~\eqref{eq:pc-small-k}). 
We note that for late and intermediate transition to the USR  the  variation  of this parameter doesn't impact our
results substantially. For early onset of the USR phase, Figure~\ref{fig:pc-dN-early} shows how the power spectrum  correction 
is affected by a change in the duration of the USR phase.

The smoothness of the transition to and from the USR epoch, captured by the parameter
$\Delta \eta$, plays an important role in dictating the amplitude of the loop-level 
contribution as $\pcq(k) \propto 1/\Delta\eta^2$.
For small $\Delta\eta$, $\pcq(k)$ could be a significant fraction of  or dominate the first-order contribution. In the intermediate onset case, $\pcq(k)$ can become as 
large as 30\% of $\ps(k)$  and it can dominate the lower order contribution for the early onset case (Figures~\ref{fig:pc-k1-interm} and~\ref{fig:pc-deta-early}). However, near-instantaneous transition is an abstraction used here to illustrate the impact of its smoothness. Any realistic inflationary  model driven by a 
potential would limit the sharpness of this transition. We will consider such models in our future work.  We note that a smoother transition also  suppresses   the oscillations  caused by a near-instantaneous transition in both the first-order power spectrum and its correction (e.g. Figure~\ref{fig:pc-k1-late}).

In the light of recent works  (e.g. \cite{Kristiano:2022maq,Kristiano:2023scm,Riotto:2023hoz,Franciolini:2023lgy}), one 
aim of this paper is to investigate the validity of perturbative expansion. Kristiano and Yokoyama \cite{Kristiano:2022maq,Kristiano:2023scm} have recently considered loop-correction to power spectrum based on  third-order interaction Hamiltonian. They concluded that USR models consistent with PBH formation result in too large a correction at large scales
to be compatible with perturbation expansion (see~\cite{Riotto:2023hoz,Fumagalli:2023hpa} for critique of this work). We consider the fourth-order interaction Hamiltonian in our work, which, as discussed above, gives a lower order correction to the power spectrum, and find that such models (late USR onset model) do not violate the validity of perturbation theory. However, we also consider intermediate 
and early USR models in our work and find that these models could lead to large corrections to the first-order power spectrum. 
We also find that the correction could be negative (we discuss this issue in detail in Appendix~\ref{app:pc-struc}) and dominate over the first-order power spectrum, which essentially indicates the necessity of contributions from higher 
order loops to restore the overall positivity of the total  power spectrum.

One encounters both infra-red and ultra violet divergences in the computation of 
loop corrections. This issue has been extensively discussed in the literature (e.g. \cite{Collins:2005nu,Weinberg:2005vy,Collins:2003mj,Collins:2003ze,
Burgess:2009bs,Gerstenlauer:2011ti,Seery:2010kh,Senatore:2009cf} and references therein). We have  already briefly  discussed  the nature of the ultra-violet divergence of  $\pcq(k)$  in both slow roll and USR models. These arise from the modes in the sub-Hubble  
regime ($k\eta \gg 1$) and can be regularized and renormalized within the framework of a more complete theory of inflationary dynamics
(e.g. for a detailed discussion see \cite{Senatore:2009cf,Weinberg:2005vy}). 
This has motivated us to neglect the sub-Hubble contribution to $\pcq(k)$ (for recent works that adopt a similar approach see e.g. \cite{Kristiano:2022maq,Franciolini:2023lgy}).
In Appendix~\ref{app:divergence}, we explore the structure of both the infra-red and 
ultra-violet divergences that arise in  the computation of two-point correlation in a generic 
inflationary model. 

An immediate extension of this analysis is the examination of models with 
realistic potentials that achieve an epoch of USR due to features such as 
inflection points in their form.  We intend to compute the $\pcq(k)$ arising 
from such models and understand their impact on the behaviour of power spectrum  (e.g. the smoothness of SR-USR-SR transitions) in terms of the features   and 
parameters of the potential. Depending on the smoothness of the SR-USR transition, more terms in the quartic action 
could  make significant contribution to  $\pcq(k)$. 
Also, the evolution of perturbations in these models may not be analytically 
tractable  and we expect  to perform this computation numerically. Finally,
our work allows us to predict  corrections to the power spectrum which could be 
probed by CMB, galaxy clustering, and Lyman-$\alpha$ data, e.g. in the intermediate onset case we find significant correction to 
the first-order power spectrum for a range of wavenumbers (Figure~\ref{fig:pc-k1-interm}). In particular, CMB data is sensitive  to  modifications larger than a few percent to the shape of the power 
spectrum.  We hope to compare our theoretical predictions with cosmological data sets  in the near future. 

\section*{Acknowledgements}

The authors would like to thank J\'{e}r\^{o}me Martin for discussions.
HVR thanks Vincent Vennin for helpful comments.
SM and LS wish to thank the Indian Institute of Technology (IIT) Madras, Chennai, 
India, for support through the Exploratory Research Project~RF22230527PHRFER008479.
HVR thanks the Raman Research Institute (RRI) for support through a postdoctoral 
fellowship.
SM and LS would like to thank RRI, and HVR and SS wish to thank IIT Madras,
respectively, for hospitality where parts of this work was carried out.


\appendix

\section{Absence of boundary term contributions}\label{app:boundary}

There have been arguments for the possibility of amelioration or cancellation of
loop contributions arising from cubic order action, by the contributions from the
boundary terms of the action of same order~\cite{Fumagalli:2023hpa,Tada:2023rgp}. 
In this appendix, we discuss the possibility for the quartic order action and show 
that such a cancellation is unlikely at this order.

Towards this, let us briefly review the argument in case of the cubic order action. 
The bulk and the temporal boundary terms relevant for the model of interest in the 
Hamiltonian density at the cubic order are
\begin{eqnarray}
H^{(3,\,\rm bulk)}_{\rm int}(\eta, \bm x) &=& -\f{1}{2}\,a^2(\eta) \epsilon_1(\eta) 
\epsilon_2'(\eta) \zeta^2(\eta, \bm x)\zeta'(\eta, \bm x)\,, \\
H^{(3,\,\rm boundary)}_{\rm int}(\eta, \bm x) &=& \f{1}{2}\,\f{\d}{\d \eta}
\bigg[a^2(\eta)\epsilon_1(\eta)\,\epsilon_2(\eta) \zeta^2(\eta, \bm x) 
\zeta'(\eta, \bm x) \bigg]\,.
\end{eqnarray}
The strong loop level contributions from these terms at the transitions is due to 
the behavior of $\epsilon_2' \propto \delta(\eta-\eta_{1,2})$, as discussed in 
the main text. The structure of the loop contribution to the power spectrum is of
the form
\bea
\langle 0 \vert \hat \zeta_{\bm k}(\ee) \hat \zeta_{{\bm k}'}(\ee) \vert 0 \rangle_{_{\rm C}} 
&=& -\left\langle\int^{\ee} \d \eta_1
[[\zeta_\vk(\ee) \zeta_{\vk'}(\ee), H^{(3)}_{\rm int}(\eta_1,\bm x)],
\int^{\eta_1} \d \eta_2\,H^{(3)}_{\rm int}(\eta_2,\bm x)]\,\right\rangle.\,\,
\eea
To understand the cancellation of contributions, we shall focus first on the 
contribution from the boundary term. When substituted in the above expression, we 
can see that the integral over $\eta_2$ is straightforward to perform, as it just 
cancels the derivative with respect to $\eta_2$. But as we perform the integral 
over $\eta_1$, there arises a term due to integration by parts, that contains 
$\epsilon_2'$ which then behaves as a Dirac delta function at the transitions. 
This contribution then mimicks the one arising from the bulk term which already 
contains $\epsilon_2'$. The relative sign difference that arises from the boundary 
contribution due to integration by parts leads to the aforementioned cancellation 
of these contributions. For explicit calculation, refer the derivation presented 
in~\cite{Fumagalli:2023hpa}. Refer also~\cite{Firouzjahi:2023bkt}, where it is 
argued that the cancellation is not exact and there still remains a non-negligible 
contribution. The exact computation of this cancellation at the cubic order is 
tangential to the aim of our work.

To investigate a similar possibility of cancellation of loop contribution at 
the quartic order, we need the set of boundary terms of the quartic order action. 
In the absence of any earlier work explicitly detailing these boundary terms, the 
computation of complete action at the quartic order is a work of its own merit. 
However, for the purpose of our analysis we assume the generic structure of a
boundary term that may possibly cancel the computed $\pcq(k)$ and argue its 
feasibility.

Let us assume that there exists a temporal boundary term at the quartic order that
leads to a Hamiltonian density of the form
\bea
H^{(4,\, \rm boundary)}_{\rm int}(\eta, \bm x) &=& \f{\d}{\d \eta} \bigg[\xi(\eta)
\zeta^3(\eta, \bm x)\zeta'(\eta, \bm x) \bigg]\,.
\eea
Recall that the combination of slow-roll parameters in the bulk term we considered
was $\epsilon_2\epsilon_3\epsilon_4\epsilon_5$ (cf.~Eq.~\ref{eq:Hint}).
To counter such a term, we need the boundary term to have 
$\xi(\eta) \simeq \epsilon_2\epsilon_3\epsilon_4$ barring numerical constants and
factors involving scale factor and potential.
With such a boundary term, the contribution to $\pcq(k)$ shall be
\bea
\pcq(k) & \simeq & \f{\xi(\ee)}{z^2(\ee)} \ps(k,\ee)\,\int \d \ln q\, \ps(q,\ee)\,.
\eea
Note that the derivative in the Hamiltonian density is cancelled by the single 
integral over time and all the quantities in this expression is evaluated close 
to the end of inflation $\ee$. At this time, the higher order slow roll parameters 
present in $\xi(\ee)$, such as $\epsilon_4 = 0$, as per our three-phase model of USR.
This is unlike the contribution from the cubic order, where the boundary term
led to contributions at the transition because of the presence of two integrals
over time.
Therefore, despite the presence of any higher order slow-roll parameters in the 
boundary terms, their contribution shall not cancel or mitigate the contribution 
from the bulk term that we have computed as $\pcq(k)$.


\section{Contractions and Feynman diagrams}\label{app:contrac}

We briefly discuss the types of contractions that arise in computing the 
loop-level contribution to the two-point correlation due to the six-point 
function arising from the Hamiltonian at the quartic order.
We also present the Feynman diagrams corresponding to these contributions.
Recall that the leading contribution to 
$\langle 0 \vert \hat \zeta_{\bm k}(\ee) \hat \zeta_{{\bm k}'}(\ee) \vert 0 \rangle$
from $\hat H^{(4)}_{\rm int}$ is
\begin{eqnarray}
\langle 0 \vert \hat \zeta_{\bm k}(\ee) \hat \zeta_{{\bm k}'}(\ee) \vert 0 \rangle_{_{\rm C}}
& = & 
- i \langle 0 \vert \left[ \hat \zeta_{\bm k}(\ee) \hat \zeta_{{\bm k}'}(\ee)\,,
\int \d \eta \,{\cal T} \left( \hat H^{(4)}_{\rm int}(\eta, \bm x) \right) \right]\vert 0 \rangle\,.
\end{eqnarray}
We expand it explicitly in terms of background quantities and $\zeta$, and obtain
\begin{eqnarray}
& = & 
\f{i}{72} \int \d \eta \, a^4 V (\epsilon_1\epsilon_2\epsilon_3\epsilon_4\epsilon_5) 
\int \d^3\bm x \, 
\langle 0 \vert \hat \zeta_{\bm k}(\ee) \hat \zeta_{{\bm k}'}(\ee)
\hat \zeta^4(\eta,\bm x) \vert 0 \rangle \nn \\
& & +~\text{hermitian conjugate}\,, \\
& = & 
\f{i}{72} \int \d \eta \, a^4 V (\epsilon_1\epsilon_2\epsilon_3\epsilon_4\epsilon_5) 
\int \f{\d^3\bm x}{(2\pi)^6} \, \int \d^3 \bm q_1 \int \d^3 \bm q_2 
\int \d^3 \bm q_3 \int \d^3 \bm q_4 \nn \\
& & \langle 0 \vert \hat \zeta_{\bm k}(\ee) \hat \zeta_{{\bm k}'}(\ee)
\hat \zeta_{\bm q_1}(\eta) \hat \zeta_{\bm q_2}(\eta) \hat \zeta_{\bm q_3}(\eta) 
\hat \zeta_{\bm q_4}(\eta)\vert 0 \rangle\,
\mathrm{e}^{i\,({\bm q}_1+{\bm q}_2+ {\bm q}_3+ {\bm q}_4) \cdot {\bm x}} \nn \\
& & +~\text{hermitian conjugate}\,.
\end{eqnarray}
There arises 2 types of contractions amongst these six operators, with a total of 
15 possible permutations of wavenumbers $q_1,\,q_2,\,q_3$ and $q_4$. 
We first identify the type of contraction that contains 3 possible permutations
and identically vanish.
These are terms that involve contraction between $f_k$ and $f_{k'}$ and two of 
modes with wavenumbers $q_1,\,q_2,\,q_3$ and $q_4$ contracting with the rest of 
the two. Such a term will have the form
\begin{eqnarray}
\langle 0 \vert \hat \zeta_{\bm k}(\ee) \hat \zeta_{{\bm k}'}(\ee) \vert 0 \rangle_{_{\rm C}}
& \supset & 3\left(\f{i}{72}\right) \delta^{(3)}(\bm k + \bm k') \vert f_k(\ee) \vert^2 
\int \d \eta \int \f{\d^3 \bm x}{(2\pi)^6} \, a^4 V
(\epsilon_1\epsilon_2\epsilon_3\epsilon_4\epsilon_5) \nn \\
& & \times \int \d^3 q_1
\vert f_{q_1}(\eta) \vert^2 \int \d^3 q_2 \vert f_{q_2}(\eta) \vert^2\,
+~\text{complex conjugate}\,.
\label{eq:pc4-vanish}
\end{eqnarray}
The factor of $3$ in the prefactor is due to the number of permutations possible
amongst the wavenumbers $q_1,\,q_2,\,q_3$ and $q_4$ for this type of contraction.
We can clearly see that the integrals involved in this expression are over purely
real quantities and hence the $i$ in the prefactor makes the whole term, a purely 
imaginary quantity that is added to its complex conjugate. Hence this type of 
contraction gives identically zero contribution to $\pcq(k)$\,.

\begin{figure}[t]
\centering
\includegraphics[width=12.5cm]{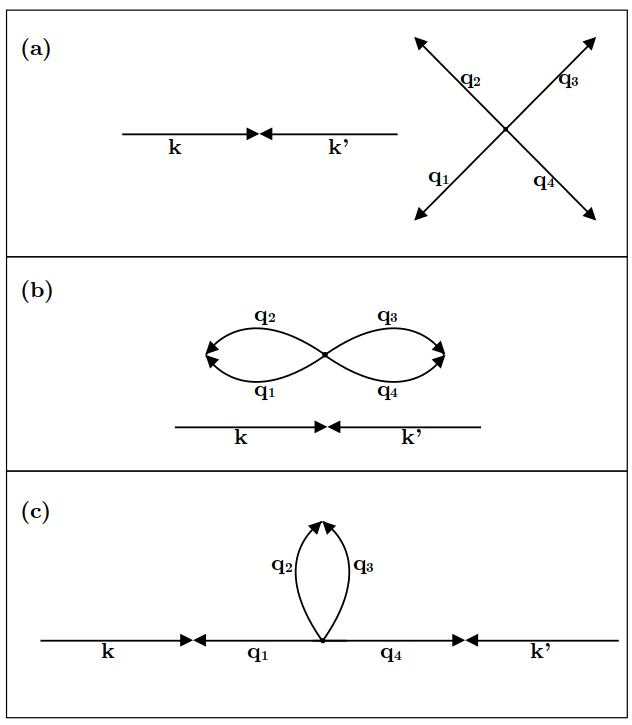}
\caption{We present the Feynman diagrams corresponding to the tree-level power
spectrum $\ps(k)$ (on left) and the interaction Hamiltonian at the quartic order
$H^{(4)}_{\rm int}$ (on right) in panel (a).
The solid lines represent the mode functions of $\zeta_k$ corresponding to the 
wavenumbers with which they are labelled and contractions are denoted by coincident
arrow heads. In panel (b), we construct the Feynman diagram corresponding to the 
vanishing loop-level contribution to $\pcq(k)$ as given in Eq.~\eqref{eq:pc4-vanish}.
Lastly, we illustrate in panel (c), the Feynman diagram corresponding to the 
non-vanishing contribution to $\pcq(k)$ as expressed in Eq.~\eqref{eq:pc4-nonvanish}.}
\label{fig:F-diagrams}
\end{figure}

To construct Feynman diagrams for such contributions, we use the diagrams
representing $\ps(k)$ and $H^{(4)}_{\rm int}$ as illustrated in the top panel of
Fig.~\ref{fig:F-diagrams}. The contraction of modes with $k$ and $k'$ and 
those with $q$ amongst themselves can then be represented as a diagram 
illustrated in the middle panel of Fig.~\ref{fig:F-diagrams}.

The other type of contraction lead to $12$ possible permutations in the
wavenumbers $q_1,\,q_2,\,q_3$ and $q_4$\,. These terms involve contraction of 
$f_k$ with one of the $f_{q_1},\,f_{q_2},\,f_{q_3}$ and $f_{q_4}$\,, and
$f_{k'}$ with another of them. The remaining two of modes are contracted between
themselves\,. This type of term will have the form
\begin{eqnarray}
\langle 0 \vert \hat \zeta_{\bm k}(\ee) \hat \zeta_{{\bm k}'}(\ee) \vert 0 \rangle_{_{\rm C}}
& \supset & 12\left(\f{i}{72}\right) \delta^{(3)}(\bm k + \bm k') f_k(\ee)\,f_{k'}(\ee)
\int \f{\d \eta}{(2\pi)^3} \, a^4 V
(\epsilon_1\epsilon_2\epsilon_3\epsilon_4\epsilon_5) \nn \\
& & \times f^\ast_{k}(\eta)\,f^\ast_{k'}(\eta) \int \d^3 q \vert f_{q}(\eta) \vert^2\,
+~\text{complex conjugate}\,.
\label{eq:pc4-nonvanish}
\end{eqnarray}
The factor of 12 arises from the number of permutations of possible contractions
between each of $k$ and $k'$ with two of $q_1,\,q_2,\,q_3$ and $q_4$\,.
This is the term that we have presented in the main text in 
Eq.~\eqref{eq:zeta-zeta-correction}\, and further analysed for the model of 
interest. The Feynman diagram corresponding to this contribution is presented in 
the bottom panel of Fig.~\ref{fig:F-diagrams}.

\section{On the divergence in the sub-Hubble regime}\label{app:divergence}

We shall briefly comment on the divergence that arises in the integral over loop 
momenta due to the sub-Hubble behaviour of the mode functions and possible 
measures to mitigate it in computation of loop-level contributions.   

The starting point of this discussion is Eq.~(\ref{eq:vk-I}), which gives the mode function
in the slow roll regime. For $k\eta \gg 1$, the mode function reduces to the solution in flat space.
It can be shown that the evolution of Mukhanov-Sasaki variable in this limit mimics the behaviour
of the fluctuation of a scalar field $\delta\phi$ in Minkowski space (e.g. \cite{2008cosm.book.....W} for a detailed 
discussion in Newtonian gauge. For $k\eta \gg 1$, $\delta\phi$ behaves as a free field in flat space.  This behaviour
is seen to be gauge-invariant by casting relevant variables in terms of   gauge invariant variables $v_k$ or $\zeta_k$, e.g. \cite{dodelson}). Therefore, the divergence arising from  the presence of $k\eta$ term  in Eq.~(\ref{eq:vk-I}) has its origin in
free scalar field theory in flat space. It is readily seen that for $k\eta \gg 1$, the dimensionless power spectrum
$k^3 |v_k|^2$ divergences as $k^2$ as $k\rightarrow \infty$.

This issue is clarified further by examining the two-point correlation of $\zeta$ in configuration 
space computed at a given $\eta$ during a standard slow roll evolution of 
inflation. It can be expressed in terms of the associated mode functions as
\bea
\langle \hat \zeta(\eta, {\bm x}) \hat \zeta(\eta,{\bm x}') \rangle &=&
\left \langle \int \f{\d^3 {\bm q}_1}{\sqrt{2\pi^3}}\,\int \f{\d^3 {\bm q}_2}{\sqrt{2\pi^3}}\,
\hat \zeta_{q_1}(\eta) \mathrm{e}^{-i {\bm q_1} \cdot {\bm x}} 
\hat \zeta_{q_2}(\eta) \mathrm{e}^{-i {\bm q_2} \cdot {\bm x}'} \right \rangle\,, \\
&=& \int \f{q^2 \d q}{2\pi^2}\,
f_{q}(\eta) f^\ast_{q}(\eta)
\f{\sin (q \vert {\bm x}- {\bm x}'\vert)}{q \vert {\bm x} - {\bm x}' \vert}\,, \\
&=& \int^\infty_{-\infty} \d \ln q \, \ps(q,\eta)\,
\f{\sin (q \vert {\bm x}- {\bm x}'\vert)}{q \vert {\bm x} - {\bm x}' \vert}\,, \\
& \simeq & \int^{k_\Delta}_{-\infty} \d \ln q \, \ps(q,\eta)\,,
\eea
where we have used the definition of $\ps(q,\eta)$ (Eq.~\eqref{eq:ps-def})
and the sharply localized nature of $\sin(x)/x$ over $x < 1$ to put an approximate 
upper limit for the integral as $k_\Delta = 1/\vert {\bm x}- {\bm x}' \vert$\,. 
Such an integral is similar to the one we encounter in the computation of loop-level 
contributions. Let us evaluate it in the typical scenario of slow roll inflation 
where the expression of $\ps(q,\eta)$ is
\begin{equation}
\ps(q,\eta) \simeq \f{H^2}{8\pi^2\Mpl^2\epsilon_1}\left( 1 + \f{k^2\eta^2}{2} \right)\,,
\end{equation}
up to corrections of ${\cal O}(\epsilon_1)$.
On performing the integral using this expression, we obtain
\bea
\langle \hat \zeta(\eta, {\bm x}) \hat \zeta(\eta,{\bm x}') \rangle &=&
\ps^0\,\bigg[ \ln \left( \f{k_\Delta}{k_{\rm min}} \right) 
+ \f{\eta^2}{2}(k_\Delta^2 - k_{\rm min}^2) \bigg]\,,
\eea
where $\ps^0 = H^2/(8\pi^2\Mpl^2\epsilon_1)$ and 
$k_\Delta = 1/\vert {\bm x}- {\bm x}' \vert$\,.
We see that the logarithmic divergence in the limit $k_{\rm min} \to 0$
and a quadratic divergence in the limit $k_\Delta \to \infty$ i.e., when we
take ${\bm x} \to {\bm x}'$, the coincident limit. There also arises a 
logarithmic divergence when $k_\Delta \to \infty$, but it is sub-dominant 
compared to the quadratic one.

It is important to note that the quadratic divergence is not an issue when 
the separation of the points ${\bm x}$ and ${\bm x}'$ is larger than the comoving 
Hubble radius, i.e. when the corresponding wavenumber $k_\Delta$ is in 
the super-Hubble regime with $k_\Delta\eta < 1$.
But it is significant when the points are deep inside the comoving Hubble radius 
i.e., when $k_\Delta$ is in the sub-Hubble regime with $k_\Delta\eta > 1$.

Focussing on the term leading to the quadratic divergence, we can recast it as
\bea
\langle \zeta(\eta, {\bm x}) \zeta(\eta,{\bm x}') \rangle & \simeq &
\f{H^2}{16\pi^2\Mpl^2\epsilon_1}k_\Delta^2\eta^2\,, \\
& \simeq & \f{L^2_{_{\rm Pl}}}{(2 \pi a\,\vert {\bm x} - {\bm x}' \vert)^2}
\f{1}{4\epsilon_1}\,,
\eea
where $L_{_{\rm Pl}} \equiv 1/\Mpl$.
This form of the divergence is similar to that of the two-point correlation of 
a free, massless scalar field in flat spacetime.   It can be easily shown, for a 
massless scalar field (say $\chi$) in a background described by Minkowski 
metric, the two-point correlation is
\bea
\langle \chi (\eta, {\bm x}) \chi (\eta,{\bm x}') \rangle & \simeq &
\f{1}{(2 \pi \,\vert {\bm x} - {\bm x}' \vert)^2}\,.
\eea
In our case, $\chi$ could be fluctuation of the scalar field, $\chi =\delta\phi$.  The same quantity, when computed in a FLRW background in the sub-Hubble regime
becomes
\bea
\langle \chi (\eta, {\bm x}) \chi (\eta,{\bm x}') \rangle & \simeq &
\f{1}{(2 \pi\, a \,\vert {\bm x} - {\bm x}' \vert)^2}\,.
\eea
So, we can see that the quadratic divergence arises when the physical distance
between the points involved in the two-point correlation vanishes. It can be due to 
setting the comoving separation $\vert {\bm x}-{\bm x}' \vert \to 0$, or if 
the background dynamics forces the points with a finite comoving separation to coincide
in physical units as $a \to 0$.

This comparison illustrates that, deep inside the Hubble radius $k\eta \gg 1$,  the divergence in $\langle \zeta(\eta, {\bm x}) 
\zeta(\eta,{\bm x}') \rangle$ is essentially due to the 
gauge-invariant variable  $\zeta$ behaving as   a massless scalar field in an inflating background (for a detailed discussion see e.g. \cite{2008cosm.book.....W}). 
Only close to the Hubble exit and for a few e-folds in the super-Hubble regime 
does the mode become sensitive to the model-dependent dynamics. The divergences of such a system 
can be removed by introducing appropriate counter-terms that modify the bare potential (e.g. see \cite{Boyanovsky:2005px,Boyanovsky:2005sh,Sloth:2006az} for details). Similar counter terms can be introduced to remove divergences in perturbative loop expansion  after appropriate regularization (e.g. \cite{Weinberg:2005vy,Senatore:2009cf}; for further details on this issue see \cite{Collins:2005nu,Collins:2003zv,Collins:2003mj,Collins:2003ze,
Burgess:2009bs,Gerstenlauer:2011ti,Seery:2010kh}).

For the sake of completeness, we give  the form of the sub-Hubble contribution to 
$\pcq(k)$. This is the   counterpart of 
Eq.~\eqref{eq:Pc_superH} that arises  from sub-Hubble modes
\bea
\pc^{(4-{\rm sub})}(k) &=& \f{i\,\epsilon_2^{\rm II}}{2k^3\eta_2\Delta\eta^2}
\left(\f{H^2}{8\pi^2\epsilon_{1f}}\right)^2 {\cal F}^2(\alpha_k,\beta_k,\ee) \nn \\ 
& & \times
\Bigg\{ \big({\cal F}^\ast(1,0,\eta_1)\big)^2\left( \f{\eta_2}{\eta_1} \right)^7 
\f{1}{2}\left(\f{k_{\rm max}}{k_1}\right)^2 \nn \\
& & - \big( {\cal F}^\ast(\alpha_k, \beta_k, \eta_2) \big)^2
\Bigg[\f{1}{2}\left(\f{k_{\rm max}}{k_2}\right)^2 
+\int_{\ln k_2}^{\ln k_{\rm max}} \d \ln q \,
\Big[2\vert \beta_q \vert^2\left( 1 + \f{q^2}{k_2^2} \right) \nn \\
& & + 2\mathfrak{Re}\Bigg(\alpha_q^\ast\beta_q \mathrm{e}^{-2i\f{q}{k_2}} 
\left( \f{q^2}{k_2^2} -1 -2i\f{q}{k_2} \right)\Bigg)
\Big]\Bigg]\Bigg\} \nn \\
& & +~\text{complex conjugate}\,.
\label{eq: Pc_subH}
\eea


\section{Structure of $\pcq(k)$}\label{app:pc-struc}

Here we investigate  the general structure of $\pcq(k)$ and its behaviour
in terms of the mode functions of perturbations. Consider the two-point 
function in configuration space, which up to linear order in $H^{(4)}_{\rm int}$
can be written as
\bea
\langle \zeta(\ee,\bm x_1) \zeta(\ee,\bm x_2) \rangle &=&
\langle \zeta(\ee,\bm x_1) \zeta(\ee,\bm x_2) \rangle
-i\,\int \d \eta \int \d^3 \bm x 
\langle[\zeta(\ee,\bm x_1) \zeta(\ee,\bm x_2), H^{(4)}_{\rm int}(\eta,\bm x)]\rangle\,.\nn \\
\eea
For simplicity we  write 
$H^{(4)}_{\rm int}(\eta,\bm x) = \lambda(\eta)\,\zeta^4(\eta,\bm x)$\,, where 
$\lambda(\eta)$ contains the background quantities such as $a,\,V$ and $\epsilon_i$'s,
and shall be of dimension $M^4$\,. Comparing with Eq.~\eqref{eq:Hint}, we can
clearly see that $\lambda = -a^4V\epsilon_1\epsilon_2\epsilon_3\epsilon_4
\epsilon_5/72$\,.
We use Wick's theorem to rewrite the six-point functions of $\zeta(\eta,x)$ in the 
second term, in terms of the two-point correlations. We  can express the reduced equation 
in terms of the mode functions $f_k(\eta)$ associated with $\zeta(\eta,\bm x)$ as
\bea
\langle \zeta(\ee,\bm x_1) \zeta(\ee,\bm x_2) \rangle &=&
\int \d \ln k\, \ps(k,\eta)\, \f{\sin(k\vert \bm x_1 - \bm x_2\vert)}
{k\vert \bm x_1 - \bm x_2\vert}\nn \\
& & -12\,i\,\int \d \eta \,\lambda(\eta) \int \d^3\bm k_1 \int \d^3\bm k_2\,
\delta^{(3)}(\bm k_1 + \bm k_2)
f_{k_1}(\ee)f_{k_2}(\ee)f^\ast_{k_1}(\eta)f^\ast_{k_2}(\eta) \nn \\
& & \times\, \mathrm{e}^{i({\bm k}_1 \cdot {\bm x}_1 + {\bm k}_2 \cdot {\bm x}_2)}
\int \f{\d^3\bm q}{(2\pi)^6}\,\vert f_q(\eta) \vert^2\,
+~\text{complex conjugate}\,.
\label{eq:twopfun}
\eea
We focus on the second term in Eq.~(\ref{eq:twopfun}) that gives the  correction
to the first-order two-point correlation. The $12$ in its prefactor arises from the permutations
of the non-vanishing contractions (Appendix~\ref{app:contrac}). We call the 
term $\langle \zeta(\ee,\bm x_2) \zeta(\ee,\bm x_2) \rangle_{\rm C}$, which is
of the form
\bea
\langle \zeta(\ee,\bm x_1) \zeta(\ee,\bm x_2) \rangle_{\rm C} &=&
-12\,i\,\int \d \eta \,\lambda(\eta) \int \d^3\bm k_1 \int \d^3\bm k_2\,
\delta^{(3)}(\bm k_1 + \bm k_2)
f_{k_1}(\ee)f_{k_2}(\ee)f^\ast_{k_1}(\eta)f^\ast_{k_2}(\eta) \nn \\
& & \times \mathrm{e}^{i({\bm k}_1 \cdot {\bm x}_1 + {\bm k}_2 \cdot {\bm x}_2)}
\int \f{\d^3\bm q}{(2\pi)^6}\,\vert f_q(\eta) \vert^2\,
+~\text{complex conjugate}\,.
\eea

To better understand the contribution of each $f_k(\eta)$, we decompose the mode functions 
into amplitudes and phases, i.e.
\bea
f_k(\eta) &=& \vert f_k(\eta) \vert \mathrm{e}^{i\,\theta}\,,
\eea
where the $\vert f_k(\eta) \vert$ is the positive definite amplitude of
the mode function at a given time and $\theta(k,\eta)$ is the associated phase.
The phase is  a function of $k$ and $\eta$, and is related to $f_k(\eta)$
as
\bea
\theta(k,\eta) &=& \tan^{-1}\Bigg[ \f{\mathfrak{Im}\left[f_k(\eta)\right]}
{\mathfrak{Re}\left[f_k(\eta)\right]} \Bigg]\,.
\eea
One can understand $\theta(k,\eta)$ by inspecting the standard mode function in
case of slow roll inflation. In such a case
\bea
\theta(k,\eta) &=& \tan^{-1}\Bigg[ \f{\sin(k\eta)+\cos(k\eta)/(k\eta)}
{\sin(k\eta)/(k\eta) - \cos(k\eta)} \Bigg]\,,
\eea 
and in the sub-Hubble and super-Hubble regimes, it reduces to
$\theta \to -k\eta$ when $\vert k\eta\vert \to \infty$ and
$\theta \to -\pi/2$ when $\vert k\eta\vert \to 0$\, respectively.
In a generic case of inflation, involving deviations from slow roll, such as USR,
$\theta$ shall be a function of the model parameters that describe the dynamics
of the epoch as well. For simplicity, we shall denote it as just $\theta(k\eta)$,
with the understanding that it depends on relevant parameters through the
mode function.

Utilising this separation of mode function into amplitude and phase, we recast
the equation of $\langle \zeta(\ee,\bm x_1) \zeta(\ee,\bm x_2) \rangle_{\rm C}$
as
\bea
\langle \zeta(\ee,\bm x_1) \zeta(\ee,\bm x_2) \rangle_{\rm C} &=&
-12\,i\,\int \d \eta \,\lambda(\eta) \int \d \ln k\, \ps(k,\ee) \vert f_k(\eta) \vert^2\,
\f{\sin (k\vert \bm x_1 - \bm x_2\vert)}{k\vert \bm x_1 - \bm x_2\vert} \nn \\
& & \times \int \d \ln q\,\ps(q,\eta)\, 
\exp\big[2i\left[\theta(k\ee)-\theta(k\eta)\right]\big] \nn \\
& & +~\text{complex conjugate}\,.
\eea
We have used the definition of $\ps(k,\eta) = k^3\,\vert f_k(\eta)\vert^2/(2\pi^2)$ 
in the above equation. We can then clearly see that the complex part of the 
terms involved in only in the exponential containing the phases evaluated at
different times and $i$ outside the integrals. Thus adding this expression with
its complex conjugate becomes simpler than before. We add them and rearrange the
order of integration to arrive at
\bea
\langle \zeta(\ee,\bm x_1) \zeta(\ee,\bm x_2) \rangle_{\rm C} &=&
12\int \d \ln k\, \ps(k,\ee) 
\f{\sin (k\vert \bm x_1 - \bm x_2\vert)}{k\vert \bm x_1 - \bm x_2\vert} \nn \\
& & \times\,2\int \d \eta\, \lambda(\eta) \vert f_k(\eta) \vert^2\,
\int \d \ln q\,\ps(q,\eta)\, \sin\big[2\left[\theta(k\ee)-\theta(k\eta)\right]\big]\,.
\eea
This expression when substituted in the complete two-point correlation gives us
\bea
\langle \zeta(\ee,\bm x_1) \zeta(\ee,\bm x_2) \rangle &=&
\int \d \ln k\, \ps(k,\ee) 
\f{\sin (k\vert \bm x_1 - \bm x_2\vert)}{k\vert \bm x_1 - \bm x_2\vert} \nn \\
& & \times\bigg[ 1 + 24\,\int \d \eta\, \lambda(\eta) \vert f_k(\eta) \vert^2\,
\int \d \ln q\,\ps(q,\eta)\, \sin\big[2\left[\theta(k\ee)-\theta(k\eta)\right]
\big]\bigg]\,. \nn \\
\label{eq:ps-expansion}
\eea
The first term inside the square braces arises from the original tree-level
spectrum and the second term evidently is the correction to it. We can now
clearly identify the correction to $\ps(k)$ from $H^{(4)}_{\rm int}$, $\pcq(k)$ 
as
\bea
\pcq(k) &=& 24\,\ps(k,\ee)\,\int \d \eta\, \lambda(\eta) \vert f_k(\eta) \vert^2\,
\int \d \ln q\,\ps(q,\eta)\, \sin\big[2\left[\theta(k\ee)-\theta(k\eta)\right]\big]\,. \nn \\
\label{eq:pc-phase}
\eea
This expression of $\pcq(k)$ can be compared against its equivalent given in 
Eq.~\eqref{eq:pc-def}.
The advantage of expressing $\pcq(k)$ in this way is that we can readily identify
the magnitude of contribution, or the envelope of oscillations, at a given $k$ from 
the term multiplying the sinusoidal term. Importantly, the sign of the 
spectrum at that $k$ is determined by computing the sinusoidal term involving
the phases.

The total power spectrum, i.e. the term in the integrand apart from 
$\sin(k\vert \bm x - \bm x'\vert)/(k\vert \bm x - \bm x'\vert)$ should be 
positive. But, if the second term $\pcq(k)$ becomes dominant over the first 
term $\ps(k,\ee)$ and turns negative, it may appear that the power spectrum is
not guaranteed to be positive. However, we may also note that if $\pcq(k)$ is 
larger than $\ps(k,\ee)$, then the series is not convergent. We 
require the higher order terms to comment on the complete power spectrum. 
Therefore, the sign of the power spectrum turning negative due to $\pcq(k)$ 
should be taken as an indication of the presence of higher order terms, those 
when accounted for, shall restore positivity.


\section{Alternative modelling of the smoothness of transitions}\label{app:smooth}
In this appendix, we discuss the implications of modelling the smoothness of
the transitions between SR and USR phases in different ways.
In the main text, we have parametrized the smoothness in terms of conformal time
$\Delta \eta$. This is motivated by characterizing the sharp behavior of $\epsilon_2'$
as a Dirac delta function around the transitions. Ref.~\cite{Martin:2011sn} contains
the original usage of such a modelling in case of Starobinsky model, which is a 
specific realization of USR models. This parameter of smoothness is a dimensional 
quantity and hence introduces an additional scale in the problem. Such a modelling
of smoothness reproduces the scalar power and bi-spectra in the 
Starobinsky model~\cite{Martin:2011sn,Martin:2014kja}.
Here, we contrast such a modelling of smoothness against potential alternatives
which may be realized in different potentials effecting a phase of USR.

\begin{figure}
\centering
\includegraphics[scale=0.25]{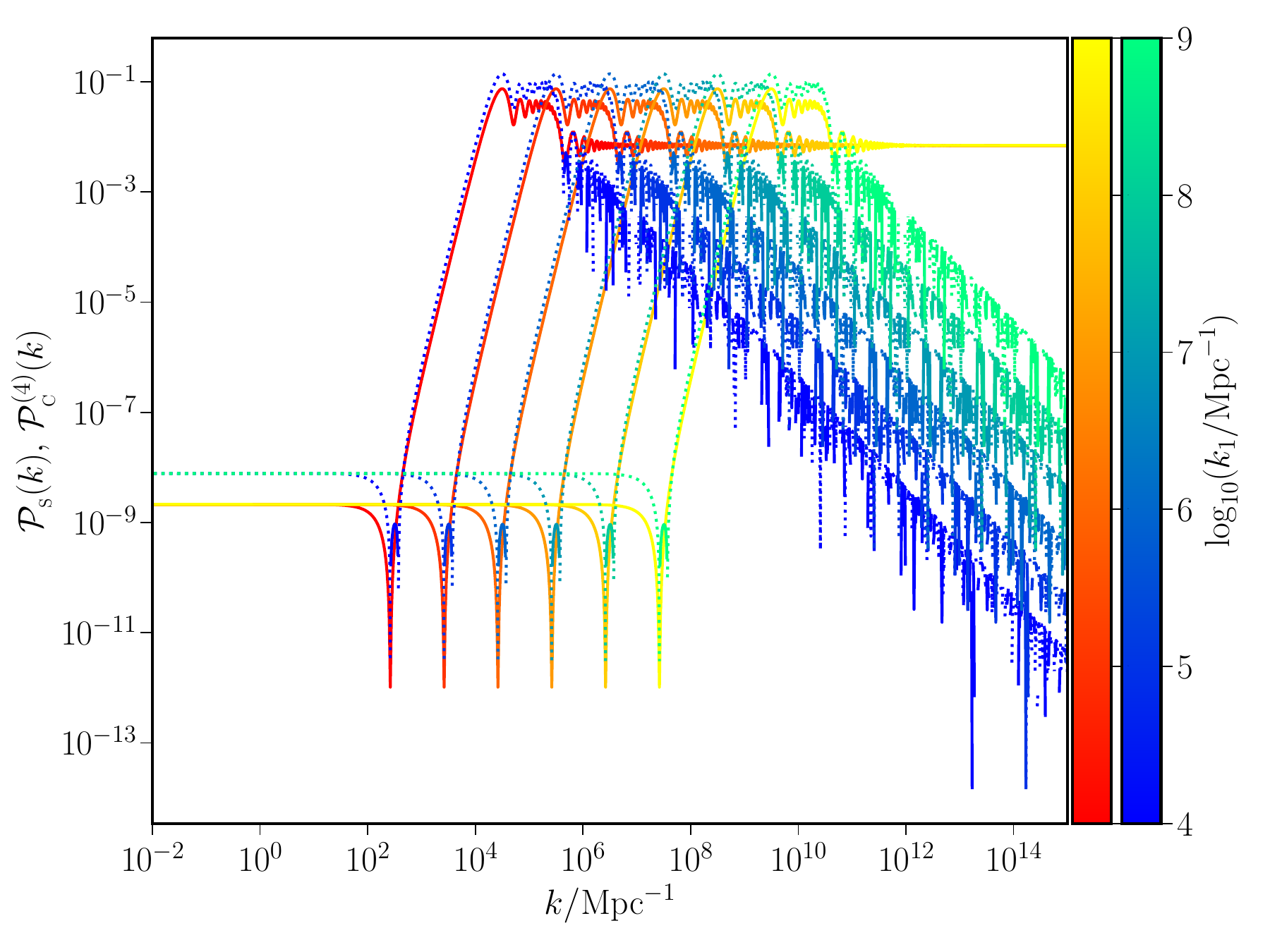}
\includegraphics[scale=0.25]{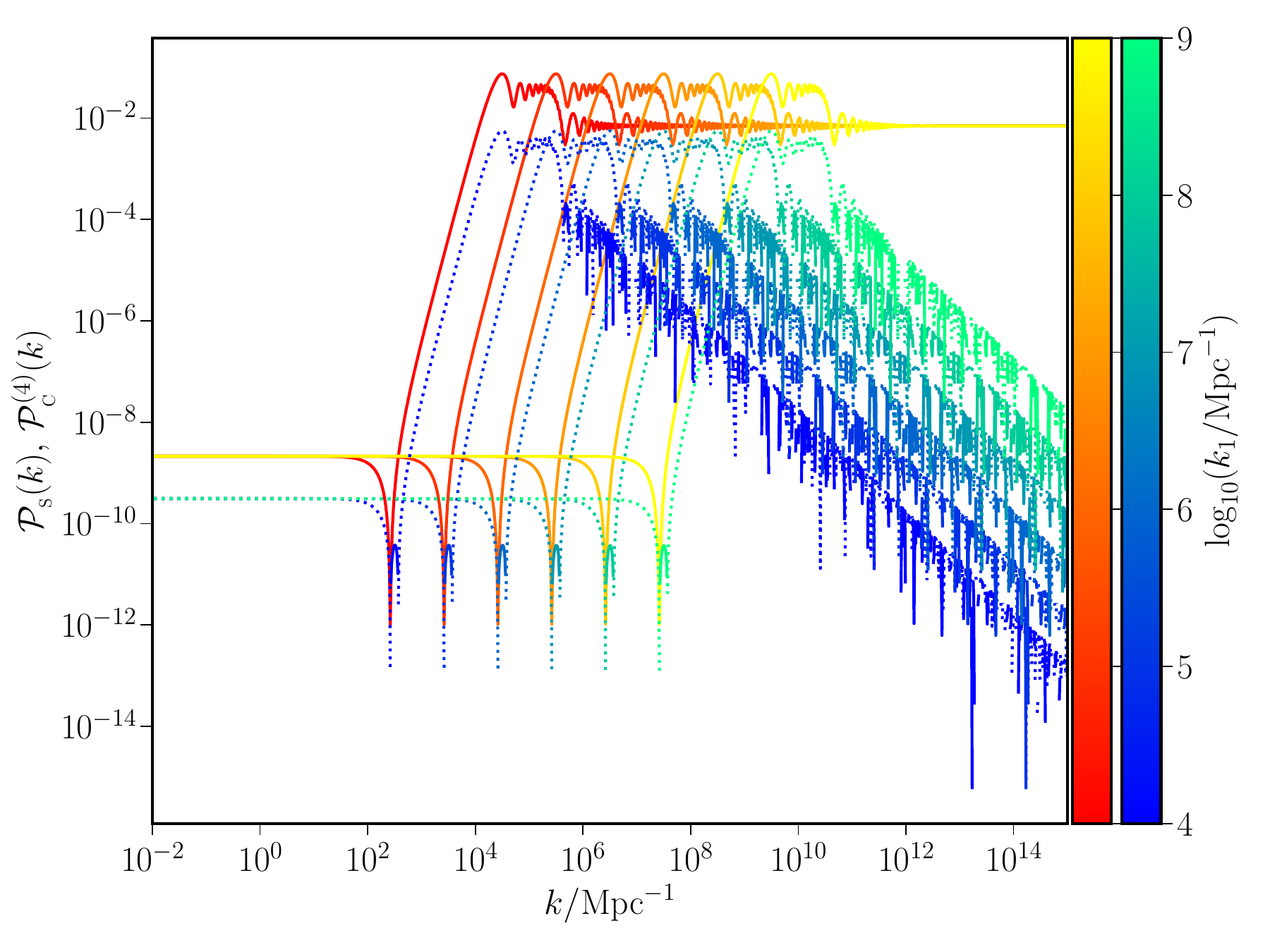}
\vskip -0.1in
\caption{The quantities $\ps(k)$ (in shades of red to yellow) and $\pcq(k)$ (in shades of 
blue to green) are plotted with $k_1$ varied from $10^4$ to $10^9\,\mpcinv$.
The smoothness of transition is parameterized as 
$\delta N = \Delta \eta_{1,2}/\eta_{1,2}$ and we set $\delta N=1/10$ (on left)
and $\delta N=5/10$ (on right).}
\label{fig:pc-dN-late}
\end{figure}

Let $\delta N \equiv \Delta \eta_{1,2}/\eta_{1,2}$, a dimensionless parameter 
characterizing the smoothness at the two transitions respectively. This parameter 
is essentially measuring the transition in terms of e-folds as
\begin{equation}
\f{\Delta \eta_{1,2}}{\eta_{1,2}} \simeq -\left[N(\eta_{1,2}+\Delta\eta) - N(\eta_{1,2})\right]=-\delta N.
\end{equation}
The overall minus sign in the RHS is due to our convention that, as inflation 
progresses, $\eta$ decreases in magnitude while $N$ increases.
But this sign difference does not play a role in the calculation. 
The crucial assumption with the definition and usage of $\delta N$ is that 
$\delta N < 1$\,.
We retain the functional form of $\epsilon_2'$ as a Dirac delta function
in terms of $\eta$. Thus we express the slow roll parameters in terms of 
$\delta N$ as
\begin{eqnarray}
\epsilon_2' &=& 
\begin{cases}
\displaystyle\f{\epsilon_2^{\rm II}}{\sqrt{\pi}\Delta\eta_1}\,
\exp\left[-\f{(\eta-\eta_1)^2}{\Delta\eta^2_1}\right]
=\displaystyle\f{1}{\sqrt{\pi}}\displaystyle\f{\epsilon_2^{\rm II}}{\eta_1\delta N}\,
\exp\left[-\f{(\eta-\eta_{1})^2}{(\eta_1\delta N)^2}\right] ~~~~~\text{around $\eta_1$}, \\
-\displaystyle\f{\epsilon_2^{\rm II}}{\sqrt{\pi}\Delta\eta_2}\,
\exp\left[-\f{(\eta-\eta_{2})^2}{\Delta\eta_2^2}\right]
=-\displaystyle\f{1}{\sqrt{\pi}} \displaystyle\f{\epsilon_2^{\rm II}}{\eta_2\delta N}\,
\exp\left[-\f{(\eta-\eta_{2})^2}{(\eta_2\delta N)^2}\right]
~~~\text{around $\eta_2$},
\end{cases}\\
\epsilon_3 &=& \mp \f{1}{\sqrt{\pi}}\f{\eta}{\eta_{1,2}\delta N}
\mathrm{e}^{-\f{(\eta-\eta_{1,2})^2}{\eta_{1,2}^2\delta N^2}},\\
\epsilon_4 &=& -\left[{1-2\f{\eta(\eta-\eta_{1,2})}{(\eta_{1,2}\delta N)^2}}\right]\,,\\
\epsilon_5 &=& 2\f{\eta(2\eta-\eta_{1,2})}{(\eta_{1,2}\delta N)^2-2\eta(\eta-\eta_{1,2})}.
\end{eqnarray}
Following the steps of calculation as outlined in the main text, we obtain
$\pcq(k)$ in terms of $\delta N$ to be
\begin{eqnarray}
\pcq(k) &=& i\,\f{\Mpl^2}{H^2}\f{\epsilon_{1_{\rm i}}
\epsilon_2^{\rm II}}{\delta N^2}\f{k^3}{2\pi^2}\,f^2_k(\ee)
\bigg[\f{[f_k^\ast(\eta_1)]^2}{\eta_1^3} \int \d \ln q\, \ps(q,\eta_1)\nn\\ 
& &- \left( \f{\eta_2}{\eta_1} \right)^6\f{[f_k^\ast(\eta_2)]^2}{\eta_2^3} 
\int \d \ln q\, \ps(q,\eta_2) \bigg] + \text{complex conjugate}\,.
\end{eqnarray}
We inspect the asymptotic behavior of $\pcq(k)$ over large scales to 
analytically understand the effect of this parametrization.
$\pcq(k)$ over large scales of $k \ll k_1 < k_2$ reduces to
\bea
\pcq(k \ll k_1) &\simeq & -\f{\epsilon_2^{\rm II}}{3\,\delta N^2}
\left(\f{H^2}{8\pi^2\Mpl^2\epsilon_{1_{\rm i}}}\right)^2
\Bigg\{ \left[ 2\l(\f{k_2}{k_1}\r)^3 -1 \right] \l[ \ln\l(\f{k_1}{k_{\rm min}}\r) + \f{1}{2} \r] \nn \\
& &-\l(\f{k_2}{k_1}\r)^6\l[\ln\l(\f{2k_2}{k_1}\r)+\gamma-\f{1}{6}\r]\Bigg\}\,.
\eea
Evidently, in the limit $\delta N \to 0$, $\pcq(k)$ diverges, which is as expected 
from the earlier results that the case of instantaneous transitions is simply 
unviable. As we fix $\delta N$ to a small, finite value and vary the onset of 
USR $\eta_1$, we find the complete behavior of $\pcq(k)$ to be as presented in 
Fig.~\ref{fig:pc-dN-late}\,.
We find that the amplitude of $\pcq(k)$ does not depend on $\eta_1=-1/k_1$, unlike
as $\pcq(k) \propto 1/k_1^2$ as seen in Figs.~\ref{fig:pc-k1-late},~\ref{fig:pc-k1-interm} and~\ref{fig:pc-k1-early}.
The features in $\pcq(k)$ in this case just get shifted similar to $\ps(k)$. 
This is because the parametrization with fixed values of $\delta N$ sets
the dimensionful smoothness $\Delta\eta_{1,2} \propto \eta_{1,2}$\,. 
Hence, in comparison to earlier result presented in say Fig.~\ref{fig:pc-k1-late},
it enhances the magnitude of $\pcq(k)$ for the cases of late onset of USR 
while it suppresses the same for early onset of USR, thereby removing the 
strong $1/k_1^2$ dependence.

Further, we see that a smooth transition of $\delta N=0.5$ leads to $\pcq(k)$ 
being $14\%$ of $\ps(k)$ and a sharper transition of $\delta N=0.1$ makes 
$\pcq(k) \geq \ps(k)$ over large scales until the peak. 
This exercise gives us an idea of how smooth the transitions have to be, in e-folds, 
for $\pcq(k)$ to be sub-dominant to $\ps(k)$.
Moreover, we find that, for the values of parameters we have worked with, 
$\pcq(k)$ is consistently negative over large scales in this parametrization of 
smoothness.
This can be understood by inspecting the asymptotic expression given above. For
our choice of parameters, the second term within the braces dominates the first.
This, along with $\epsilon_2^{\rm II}=-6$, makes the overall sign of $\pcq(k)$
negative. This is in contrast to Eq.~\eqref{eq:pc-large-k}, where the first term
dominated the magnitude.
\begin{figure}
\centering
\includegraphics[scale=0.25]{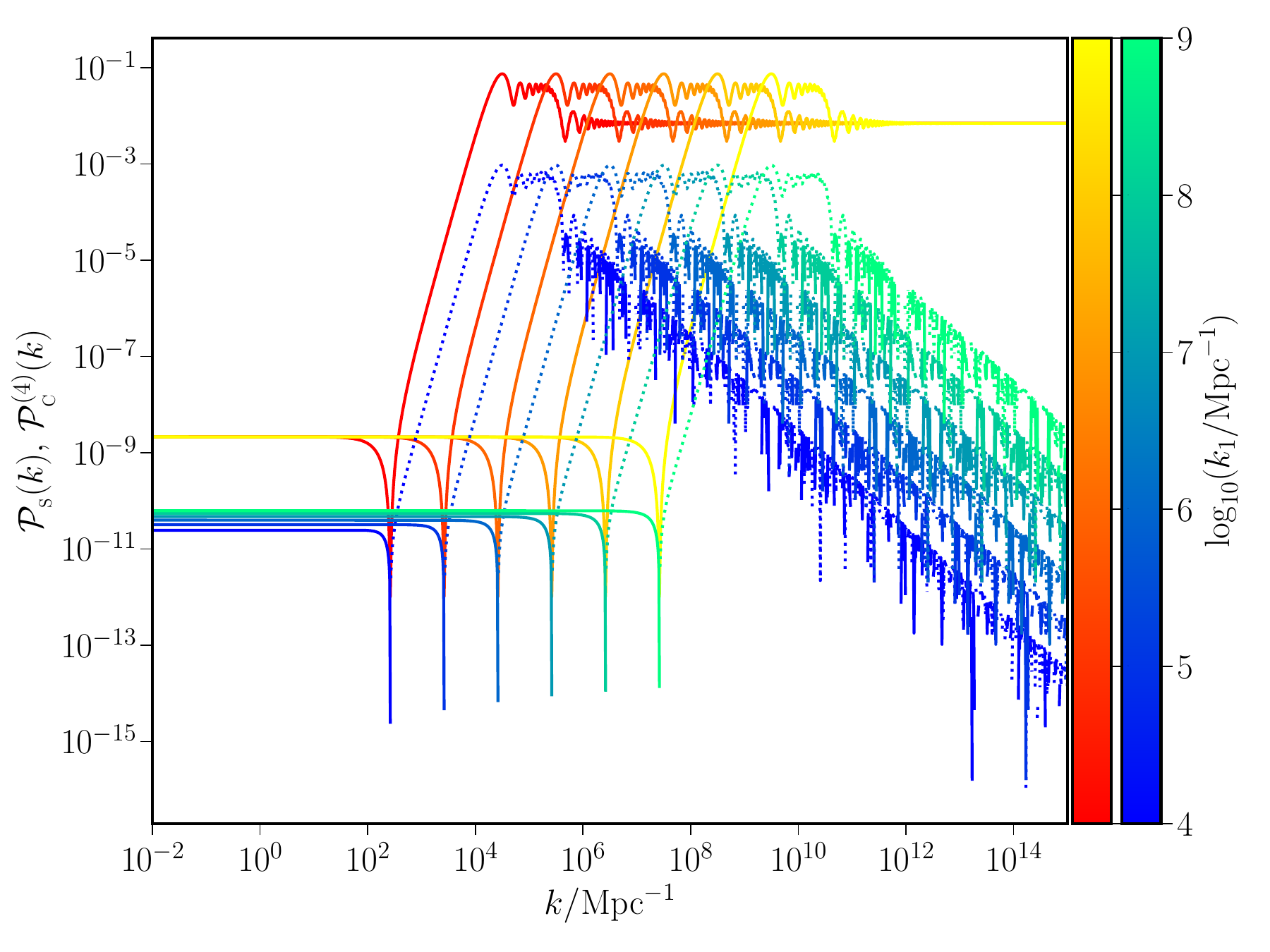}
\includegraphics[scale=0.25]{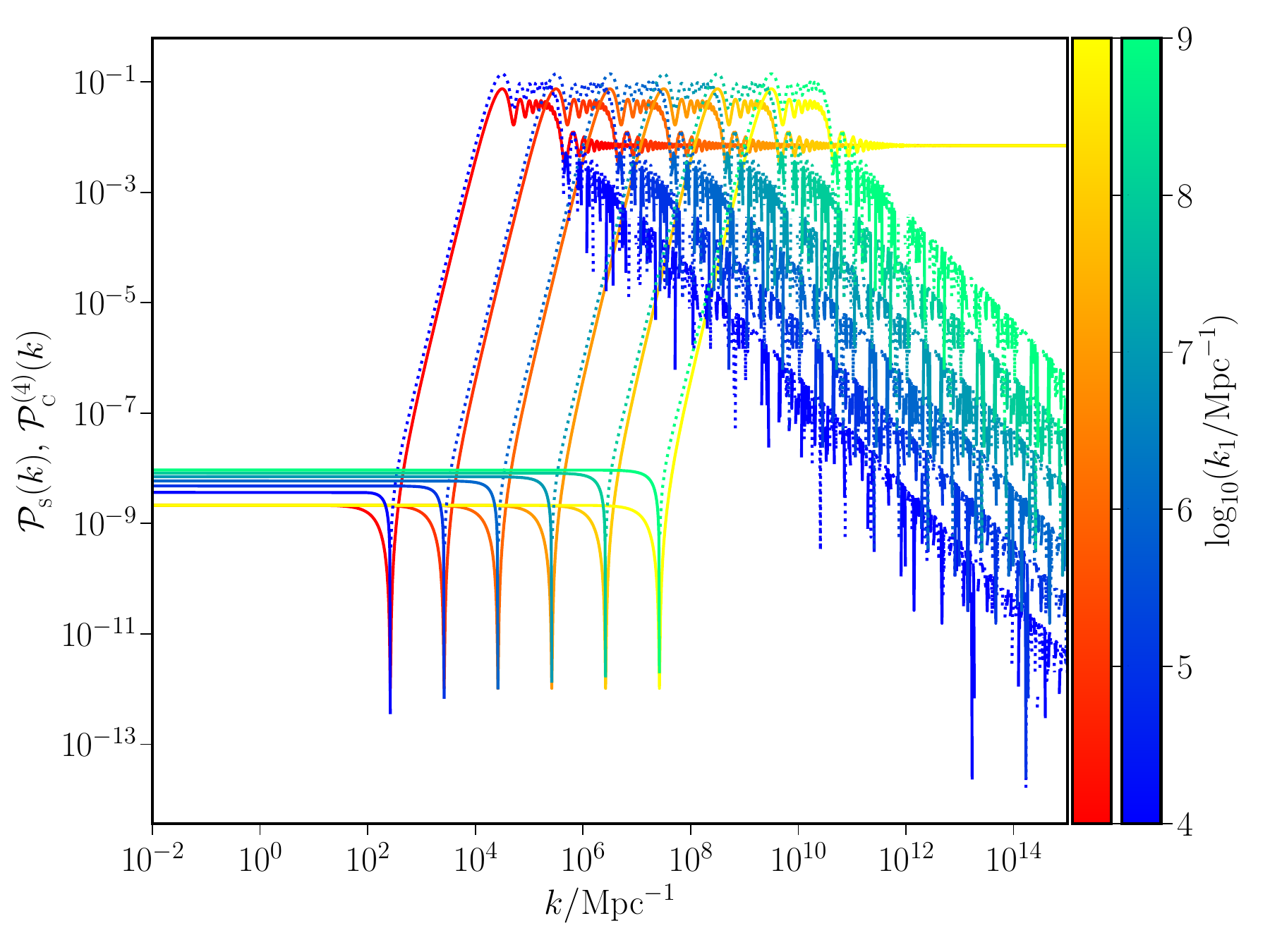}
\vskip -0.1in
\caption{The quantities $\ps(k)$ (in shades of red to yellow) and $\pcq(k)$ (in shades of 
blue to green) are plotted with $k_1$ varied from $10^4$ to $10^9\,\mpcinv$.
The smoothness of transition is parameterized as $\delta N = \Delta \eta/\eta_1$
(on left) and  $\delta N = \Delta \eta/\eta_2$ (on right).
We choose the value of $\delta N=1/10$. The amplitude of $\pcq(k)$ over large 
scale depends on $k_1$ as $\ln(k_1/k_{\rm min})$ 
[cf. Eq.~\eqref{eq:pc-large-k}].}
\label{fig:pc-dN12-late}
\end{figure}

We also present two sub-cases of the above parametrization of smoothness in terms 
of $\delta N$. We set $\Delta \eta_1=\Delta \eta_2=\Delta \eta$ in these two sub-cases. 
First, we set smoothness to be characterized as $\delta N=\Delta \eta/\eta_1$ at both 
the transitions. This implies that the second transition is smoother that the first with
\begin{equation}
\f{\Delta \eta}{\eta_2} = \f{\eta_1}{\eta_2}\delta N\,.
\end{equation}
Recall that the ratio $\eta_1/\eta_2 \gg 1$\,.
In this case, the expression of $\pcq(k)$ over large scales becomes
\bea
\pcq(k \ll k_1) &\simeq & -\f{\epsilon_2^{\rm II}}{3\,\delta N^2}
\left(\f{H^2}{8\pi^2\Mpl^2\epsilon_{1_{\rm i}}}\right)^2
\Bigg\{ \left[ 2\l(\f{k_2}{k_1}\r)^3 -1 \right] \l[ \ln\l(\f{k_1}{k_{\rm min}}\r) + \f{1}{2} \r] \nn \\
& &-\l(\f{k_2}{k_1}\r)^4\l[\ln\l(\f{2k_2}{k_1}\r)+\gamma-\f{1}{6}\r]\Bigg\}\,. 
\label{eq:pc-large-k-deln1}
\eea
The behavior of $\pcq(k)$ in this case as a function of onset of USR is presented 
in left panel of Fig.~\ref{fig:pc-dN12-late}\, with the value of $\delta N=0.1$. 
We observe that $\pcq(k)$ is sub-dominant to $\ps(k)$ throughout the range of 
scales even with this sharper value. It is due to the smoother transition at 
$\eta_2$ contributing much lesser than that at $\eta_1$.
This means that the second term in asymptotic expression above is subdominant, 
similar to Eq.~\eqref{eq:pc-large-k}\,. This leads to $\pcq(k)$ remain positive 
over large scales and have a mild dependence on $k_1$ as $\ln(k_1/k_{\rm min})$.

For the second sub-case, we set $\delta N=\Delta \eta/\eta_2$ at both the 
transitions, which makes the first transition sharper than the second with
\begin{equation}
\f{\Delta \eta}{\eta_1} = \f{\eta_2}{\eta_1}\delta N\,,
\end{equation}
and the ratio $\eta_2/\eta_1 \ll 1$\,.
In this case, the expression of $\pcq(k)$ over large scales becomes
\bea
\pcq(k \ll k_1) &\simeq & -\f{\epsilon_2^{\rm II}}{3\,\delta N^2}
\left(\f{H^2}{8\pi^2\Mpl^2\epsilon_{1_{\rm i}}}\right)^2\l(\f{k_2}{k_1}\r)^2
\Bigg\{\left[ 2\l(\f{k_2}{k_1}\r)^3 -1 \right] \l[ \ln\l(\f{k_1}{k_{\rm min}}\r) 
+ \f{1}{2} \r] \nn \\
& &-\l(\f{k_2}{k_1}\r)^4\l[\ln\l(\f{2k_2}{k_1}\r)+\gamma-\f{1}{6}\r]\Bigg\}\,.
\eea
Once again we see that the first term dominates over the second and $\pcq(k)$ has the 
mild dependence on $k_1$ as $\ln(k_1/k_{\rm min})$. In fact, the expression is very
similar to Eq.~\eqref{eq:pc-large-k-deln1}, except for the factor of $(k_2/k_1)^2$ in
the prefactor, which enhances the overall magnitude.
The complete behavior of $\pcq(k)$ in this case is presented in the right panel of 
Fig.~\ref{fig:pc-dN12-late}\,. For the same value of $\delta N=0.1$, we see that the 
overall amplitude has grown to be much larger than the previous case, becoming even 
dominant over $\ps(k)$ over large scales until the peak, as expected from the overall
factor of $(k_2/k_1)^2$ in the above expression.

This exercise with different parametrizations of smoothness gives us an insight
about the effect of the choice of smoothness on $\pcq(k)$\,.
In the absence of a realistic potential for the model, parametrization of smoothness
in terms of $\Delta \eta$ is equally valid to a parametrization in terms of 
$\delta N$.
The major takeaway about the dependence of $\pcq(k)$ on smoothness is that 
$\pcq(k) \propto 1/\Delta \eta^2 \propto 1/\delta N^2$ and hence the divergence in
the limit of instantaneous transitions.
The exact dependence of $\pcq(k)$ on $k_1$ in a model described by a potential 
shall be determined by specific features in the potential that effects the epoch of
USR.

\bibliographystyle{JHEP}
\bibliography{loops_usr}

\end{document}